\documentclass[12pt]{article}
\usepackage{a4wide}
\usepackage{graphicx}
\usepackage{dcolumn}
\usepackage{bm}
\usepackage{hyperref}
\usepackage{color}
\usepackage[normalem]{ulem}
\usepackage{subfig}
\usepackage{mathtools}
\usepackage{tikz}
\usepackage{tikz}
\usetikzlibrary{arrows}
\usetikzlibrary{decorations.pathmorphing}

\def\square{\kern1pt\vbox{\hrule height 1.2pt\hbox{\vrule width 1.2pt\hskip 3pt
   \vbox{\vskip 6pt}\hskip 3pt\vrule width 0.6pt}\hrule height 0.6pt}\kern1pt}

\newcommand{\be}{\begin{equation}}
\newcommand{\ee}{\end{equation}}
\newcommand{\bea}{\begin{eqnarray}}
\newcommand{\eea}{\end{eqnarray}}

\newcommand{\nn}{\nonumber}
\newcommand{\vv}[1]{\mathbf{#1}}

\begin{document}

{\renewcommand{\thefootnote}{\fnsymbol{footnote}}
\hfill  IGC--16/11--1\\
\medskip
\begin{center}
{\LARGE  Minisuperspace models of discrete systems }\\
\vspace{1.5em}
Bekir Bayta\c{s}\footnote{e-mail address: {\tt bub188@psu.edu}}
and Martin Bojowald\footnote{e-mail address: {\tt bojowald@gravity.psu.edu}}
\\
\vspace{0.5em}
Institute for Gravitation and the Cosmos,\\
The Pennsylvania State
University,\\
104 Davey Lab, University Park, PA 16802, USA\\
\vspace{1.5em}
\end{center}
}

\setcounter{footnote}{0}

\begin{abstract}
  A discrete quantum spin system is presented in which several modern methods
  of canonical quantum gravity can be tested with promising results. In
  particular, features of interacting dynamics are analyzed with an emphasis
  on homogeneous configurations and the dynamical building-up and stability of
  long-range correlations. Different types of homogeneous minisuperspace
  models are introduced for the system, including one based on condensate
  states, and shown to capture different aspects of the discrete system. They
  are evaluated with effective methods and by means of continuum limits,
  showing good agreement with operator calculations whenever the latter are
  available. As a possibly quite general result, it is concluded that an
  analysis of the building-up of long-range correlations in discrete systems
  requires non-perturbative solutions of the dynamical equations. Some
  questions related to stability can be analyzed perturbatively, but suggest
  that matter couplings may be relevant for this question in the context of
  quantum cosmology.
\end{abstract}

\section{Introduction}

Minisuperspace models of quantum-field theories, in particular quantum
gravity, are usually constructed by quantizing a set of configurations
obtained from the full classical theory by imposing spatial homogeneity. While
homogeneous configurations are exact (though special) solutions of the
classical theory, for various reasons they are not expected to be exact
solutions of the full quantum-field theory. For instance, uncertainty
relations would prevent both the amplitude and momentum of an inhomogeneous
mode from having zero quantum fluctuations. In an interacting theory,
fluctuations couple to expectation values, and non-zero fluctuations usually
imply that the mode expectation values cannot remain zero in time. An exactly
homogeneous (non-vacuum) solution therefore cannot be realized in a
quantum-field theory. The question of what kind of an approximation to the
full quantum theory a minisuperspace model may provide has remained open, but
recently canonical effective methods have shed some light on this question for
scalar quantum-field theories on a flat background space-time \cite{MiniSup}.

The main application of minisuperspace models is in the context of quantum
gravity, where space-time is no longer a background but quantized as
well. Several approaches to quantum gravity suggest that space or space-time
may no longer be continuous in this setting. (See for instance
\cite{OUP,Oriti}.) Discrete space may present a further obstacle to finding
exact or approximate homogeneous solutions of the theory: even if we disregard
quantum fluctuations or their back-reaction on expectation values, local moves
in a discrete structure do not respect homogeneity. At most, a coarse-grained
model which collects the accumulated action of many local moves in a single
evolution step could lead to approximate homogeneous solutions. However,
coarse-graining remains incompletely understood in discrete approaches to
background-independent gravity. (See for instance
\cite{Nets,CoarseGrainingFlow} for recent realizations.)

In order to probe these questions, we introduce here a discrete quantum system
which exhibits several interesting aspects regarding minisuperspace
models. Starting from the discrete quantum theory rather than a classical
continuum theory allows us to analyze how different features of the
interacting dynamics can be captured in simpler systems. As is well known, a
discrete theory can give rise to different continuum limits. Each of them
would then lead to a different minisuperspace model. The same result can be
seen directly by minisuperspace constructions performed for the discrete
quantum theory.

We will also analyze the discrete quantum theory in qualitative terms. In
particular, we are interested in the question of how long-range correlations
can build up in a fundamental theory and under which conditions they are
stable. If such correlations can be achieved, it is at least possible that
nearly homogeneous configurations can be the result of evolution in the
theory, rather than just of specific initial choices as implicitly made in
minisuperspace constructions. Of course, homogeneous configurations require
long-range correlations of a very specific kind which is more difficult to
analyze for a generic interacting theory. But the building-up of some kind of
long-range correlations is a pre-requisite for near homogeneity, and it can be
studied in our model in qualitative terms. The stability question will lead us
back to the ground-state configurations discussed for the various
minisuperspace models introduced here. An interesting interplay between the
full discrete theory and the models is important for the physical
interpretation of minisuperspace results.

\section{The model}

In the absence of a consistent canonical quantum theory of gravity, it is not
clear what Hamiltonian one should use to model its discrete dynamics. (See for
instance \cite{QSDI,AQGI,TwoPlusOneDef,AnoFreeWeak,OffShell} for some issues
involved in such a construction.)  In an attempt to construct tractable
models, we focus here on some of the ingredients that seem to be rather
general. We do not intend to capture the precise dynamics of quantum gravity
but rather plan to explore some properties of possible candidates for
fundamental degrees of freedom.

Several proposals of quantum gravity, going back to \cite{PenroseSpinNetwork},
are based on mathematical versions of angular momentum or spin as a
fundamental degrees of freedom. Not only spin eigenvalues but also their
proposed arrangement on a graph in space, or a spin network, are
discrete. Certain invariant combinations of spin quantum numbers can then be
defined as discrete analogs of the common continuum expressions of geometrical
measures. Moreover, spin-spin interactions can be used to introduce possible
Hamiltonians.

\begin{figure}[htbp]
\begin{center}
 \includegraphics[scale=1]{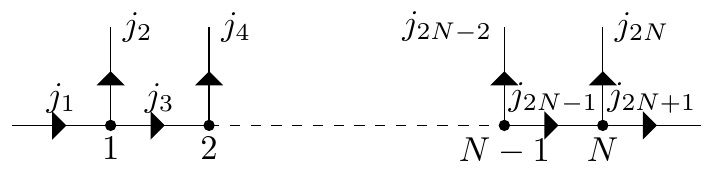}
\end{center}
\caption{The inhomogeneous one-dimensional graph $\Gamma$ with $N$ vertices
  and $2N+1$ links. The Hamiltonian (\ref{H}) is invariant under a mapping of
  spins that corresponds to a reflection of the graph in a horizontal
  direction. If the orientation of a link, indicated by an arrow, changes
  under this reflection, the corresponding mapped spin has a negative sign.
  \label{Fig:Graph}}
\end{figure}

Handling the arrangement of spins on an arbitrary graph in three spatial
dimensions can be a difficult combinatorial problem. The first simplification
we will use is a dimensional reduction: We will consider only one spatial
dimension, which could be thought of as the radial direction measuring the
distance from a non-rotating and spherical star or black hole, or more
generally a so-called midisuperspace model of general relativity. Aligned in
this direction are then several different types of spins, which roughly
correspond to independent components of the spatial metric. We arrive at a
1-dimensional graph model as illustrated in Fig.~\ref{Fig:Graph}. Here, spins
on links in the horizontal direction, called ``horizontal spins'' in what
follows, would then have a geometrical interpretation distinct from that of
``vertical spins'' on upward and open-ended links. However, such a geometrical
interpretation will not be relevant for our analysis of the interacting
dynamics. 

Specific versions of such quantum midisuperspace models with explicit
Hamiltonians have been constructed for spherically symmetric models
\cite{SphSymm,SphSymmHam} and certain types of gravitational waves
\cite{LoopCorr,EinsteinRosenQuant}. In tractable versions, one makes use of a
further reduction of the group SU(2) to the Abelian U(1). In order to have
interesting spin-spin interactions, we will not make use of this reduction
here. However, we will simplify the combinatorics by working with a single
spin on each link, instead of distinguishing between left- and right-invariant
vector fields on SU(2) as would be done in a full spin network. 

The dynamics on a spin system, such as the one illustrated in
Fig.~\ref{Fig:Graph}, is in general spin-changing as well as
graph-changing if it comes from a generic proposal of canonical
quantum gravity. That is, the Hamiltonian can contain terms that
change the irreducible representation of SU(2) on each link of the
graph, as well as terms that can create new vertices and corresponding
links of the graph. Such a dynamics is hard to control, and therefore
we assume a simplified version in which no spin-changing or
graph-changing terms occur.  Therefore, for given irreducible
representations and a fixed graph, only spin-spin interactions are
present in the Hamiltonian. We consider only local (next-neighbor)
pairwise interactions and require a certain reflection symmetry as
indicated in Fig.~\ref{Fig:Graph} and spelled out in the explicit
construction that follows.

As the discrete theory, we introduce a spin system which for a given integer
$N$ has $2N+1$ interacting spins $J_i$, $i=1,\ldots, 2N+1$. We define the
dynamics in canonical form, generated by the Hamiltonian
\begin{equation}
\label{H}
\hat{\rm{H}}_{\Gamma} = \alpha \sum \limits_{i=1}^{N} \, \big(-\hat{\vv
  J}_{2i-1} \cdot \hat{\vv J}_{2i} + \hat{\vv J}_{2i-1} \cdot \hat{\vv
  J}_{2i+1} + \hat{\vv J}_{2i} \cdot \hat{\vv J}_{2i+1}\big) 
\end{equation}
with a coupling constant $\alpha$. This operator is invariant under global
rotations of the spins: the sum of all horizontal spins,
\begin{equation} \label{G}
 \vv G:=\sum_{i=1}^{N+1} \hat{\vv J}_{2i-1}\,,
\end{equation}
commutes with the Hamiltonian. This conserved quantity can be used in some
cases to simplify equations of motion, as in Sec.~\ref{s:Small}.  We have
chosen the signs of individual coefficients of the spin products so as to make
the Hamiltonian reflection symmetric under the operation $\hat{\vv
  J}_{2i-1}\mapsto -\hat{\vv J}_{2N-2i+3}$ while $\hat{\vv J}_{2i}\mapsto
\hat{\vv J}_{2N-2i+2}$ (or $i\mapsto N-i+1$). These properties can be
illustrated by the graph model presented in Fig.~\ref{Fig:Graph}. The arrows
indicate the sign in the reflection symmetry.



In 1-dimensional models of gravity, it is often convenient to impose
polarization conditions which eliminate one of the metric components as an
independent field. In our discrete model, such a condition would then relate
the different types of spins (horizontal and vertical) to each other. Our
polarization condition used here is a constraint that corresponds to the
classical conditions
\begin{equation} \label{constraintoperator}
\vv C_i  = - \vv J_{2i-1} + \vv J_{2i} + \vv J_{2i+1}
= 0 \quad,\quad i=1,\ldots,N \,.
\end{equation}
They can be used to eliminate the vertical spins.
Also this system of constraints has coefficients chosen
so as to make it reflection symmetric: $\vv C_i\mapsto \vv C_{N-i+1}$.

We have a system of constraints in a non-symplectic Poisson manifold with
coordinates given by spin components $J_i^a$, such that standard
classifications of first or second class constraints are not available
\cite{brackets}. It is, however, straightforward to see that the constraints
do not all (Poisson) commute with one another, nor with the Hamiltonian.  The
non-zero Poisson brackets are
$\{C_i^a,C_i^b\}=\epsilon^{abc}(J_{2i-1}^c+J_{2i}^c+J_{2i+1}^c)\approx
2\epsilon^{abc} J_{2i-1}^c$ (the weak equality $\approx$ indicating that the
constraints have been used) and $\{C_i^a,C_{i+1}^b\} = -\epsilon^{abc}
J_{2i+1}^c$ for $a\not=b$, while all other components of the constraints
commute.  We are only interested in imposing the constraints as a reduction of
vertical degrees of freedom. The constraint surface remains well-defined if
the reduction constraints are imposed strongly. In particular, we can solve
the constraints so as to eliminate all vertical spins $\vv J_{2i}$ (or
$\hat{\vv J}_{2i}$), and use standard Poisson brackets (or commutators) for
the remaining $\vv J_{2i-1}$ (or $\hat{\vv J}_{2i-1}$).  All our derivations
exclusively use Poisson brackets or commutators, and therefore the model is
sufficient as a non-symplectic Poisson system. 

We use the system of partially non-commuting constraints as an example of
reduction, better known from the context of symmetry reduction. In fact, if we
combine the constraints $\vv C_i$ with additional constraints that set all the
vertical spins equal to zero, the reduction imposes homogeneity: it requires
that the remaining, horizontal spins are all equal, $\vv J_{2i-1}=\vv
J_{2i+1}$ for all $i=1,\ldots,N$. If we impose only $\vv C_i=0$ without
restricting vertical spins, we can solve for the vertical spins and obtain a
single 1-dimensional spin chain closely related to the next-neighbor
Heisenberg spin chain. The fact that the quantized constraints do not commute
with $\hat{H}$ allows us to probe for potential effects of local discrete
moves not respecting reduction constraints.

An important question in classical symmetry reduction is whether variation
commutes with reduction.  It is not always guaranteed that equations of motion
of the reduced system (extrema of the reduced action) agree with the field
equations of the full theory restricted to fields that obey the reduction
condition. Certain general conditions are known that guarantee this
commutation property (symmetric criticality), formulated mainly as conditions
on properties of the corresponding symmetry group \cite{midisup,midisup2}. In
our case, we have a reduction constraint which shares with minisuperspace
reductions the feature that it is (partially) non-commuting, but it does not
directly correspond to a symmetry group. Moreover, we are working exclusively
with Hamiltonians rather than action principles, and we do not have a
symplectic phase space.

However, instead of using general conditions on symmetry groups, it is not
difficult to test the commutation property directly. We have the reduced
Hamiltonian
\begin{eqnarray}
\label{Hreduced}
\hat{\rm{H}}_{\mathrm{red}} &=& \alpha \sum \limits_{i=1}^{N} \,
\big(- \hat{J}_{2i-1}^2 + 3 \, \hat{\vv J}_{2i-1} \cdot \hat{\vv J}_{2i+1} -
\hat{J}_{2i+1}^2 \big) \\
&=& \alpha\left(-\hat{J}_1^2- 2\sum_{i=2}^N \hat{J}_{2i-1}^2-
  \hat{J}_{2N+1}^2+ 3\sum_{i=1}^N \hat{\vv J}_{2i-1}\cdot \hat{\vv
    J}_{2i+1}\right)\,,
\label{Hreduced2}
\end{eqnarray}
which generates Heisenberg equations of motion
\begin{equation}
\label{expect5X}
\frac{{\rm d}\hat{J}_{2i+1}^a}{{\rm d} t} =  3 \, \alpha \,
\epsilon^{abc} \, \big(\hat{J}_{2i-1}^{b} \hat{J}_{2i+1}^{c} +
 \hat{J}_{2i+1}^{c} \hat{J}_{2i+3}^{b}\big) \,.
\end{equation}
The full equations for horizontal spins are
\begin{equation}
\frac{{\rm d}\hat{J}_{2i+1}^{a}}{{\rm d} t} = \alpha \, \epsilon^{abc}
\, \big(- \hat{J}_{2i+1}^c \hat{J}_{2i+2}^{b}   + \hat{J}_{2i-1}^{b}
\hat{J}_{2i+1}^{c}  + \hat{J}_{2i+1}^{c}
\hat{J}_{2i+3}^{b} + \hat{J}_{2i}^{b}  \hat{J}_{2i+1}^{c} \big)\,,  
\end{equation}
coupled to vertical spins $\hat{\vv J}_{2i}$. If we use the constraint in
order to eliminate the vertical spins in the equation of motion, we obtain
\begin{equation}
\label{expect3X}
\frac{{\rm d}\hat{J}_{2i+1}^{a}}{{\rm d} t} =  2 \, \alpha \,
\epsilon^{abc} \, \big(\hat{J}_{2i-1}^{b} \hat{J}_{2i+1}^{c} +
\hat{J}_{2i+1}^{c}\hat{J}_{2i+3}^{b}
 \big)
\end{equation}
which are not identical with the equations generated by the reduced
Hamiltonian. However, the difference is merely a constant numerical factor of
the time derivatives. 
We have interpreted the constraint imposed here as a polarization
condition. The preceding calculations have shown that there is a small
difference between imposing the polarization condition before or after
deriving equations of motion. The coupling of modes is therefore slightly
different if it is described by a reduced Hamiltonian, compared with the full
Hamiltonian on whose equations of motion the same condition would be
imposed. One can account for the difference by a simple rescaling (or a
classical renormalization) of the coupling constant, using $\frac{2}{3}\alpha$
instead of $\alpha$ in the reduced Hamiltonian.

In what follows, we will, for simplicity, work mainly with reduced
Hamiltonians. (The conserved quantity (\ref{G}) now commutes strongly
with the Hamiltonian.)  We will compare different versions of
homogeneous minisuperspace models and effective continuum theories.

\section{Minisuperspace models}

All spins in the model are coupled. It might therefore be possible
that long-distance correlations build up over time, which could be of
classical or quantum nature. A minisuperspace configuration would be
one example of a classically correlated system. There is a difference
between such a minisuperspace configuration and a homogeneous
configuration as it might be realized as a ground state of the
unreduced system because all spins would have to be identical as
degrees of freedom, not just equal as values assigned to different
links of the graph. In this subsection, we explore several questions
related to this conceptual difference. We will also see that there is
some freedom in defining different minisuperspace models, and that
selecting a model that gives predictions close to the discrete theory
requires knowledge of solutions of the discrete theory.

\subsection{A minimal minisuperspace model}

The smallest number of minisuperspace degrees of freedom can be realized by
identifying all horizontal spins on the links. Setting
$\hat{\vv J}_{2i-1}=\hat{\vv J}_{2i+1}=: \hat{\vv J}$ in the reduced Hamiltonian
(\ref{Hreduced}) gives us the minisuperspace Hamiltonian
\begin{equation}
 \hat{H}_{\rm mini}^{(1)} = \alpha N \hat{J}^2 \,.
\end{equation}
We obtain the same result if we start with the unreduced Hamiltonian and set
vertical spins equal to zero.  Clearly, this Hamiltonian commutes with all
minisuperspace degrees of freedom, given by the three components of a single
$\hat{\vv J}$. The resulting trivial dynamics is very different from the
coupled equations of the inhomogeneous theory.

\subsection{A condensate model}

Alternatively, homogeneity can be imposed at the level of states by working
with condensate states of the full theory, given by
\begin{equation} \label{Condensate}
 \Psi = \bigotimes_{i=1}^{N+1} \chi
\end{equation}
where $\chi$ is a state in the Hilbert space of a single horizontal spin. The
same individual state is therefore assumed for all links, but unlike in the
minisuperspace model, the spins remain as independent degrees of freedom. This
procedure is well known from the description of Bose--Einstein condensates,
where it results in the non-linear Gross-Pitaevsky equation for the analog of
$\chi$. In quantum gravity, the procedure has been used in particular in the
context of group-field theory
\cite{GFTCosmo,GFTCosmo2,GFTCosmo3,GFTLattice,GFTPerturb,GFTLQC,GFTCyclic,GFTImpact}, and it can
also be seen in certain approximations that go beyond minisuperspace models by
including perturbative inhomogeneity at an effective quantum level
\cite{NonLinLQC}.

A non-linear equation for the single-spin wave function $\chi$ can be derived
by first computing an effective Hamiltonian $\langle\Psi|\hat{H}|\Psi\rangle$
for (\ref{Hreduced2}) in a state of the form (\ref{Condensate}):
\begin{equation}
 \langle\Psi|\hat{H}|\Psi\rangle = -2\alpha N
 \langle\chi|\hat{J}^2|\chi\rangle + 3\alpha N \langle\chi|\hat{\vv
   J}|\chi\rangle^2 
\end{equation}
where we have identified $\langle\chi|\hat{\vv J}_i|\chi\rangle=
\langle\chi|\hat{\vv J}_j|\chi\rangle$ for $i\not=j$. Effective equations of
motion or variational ground states can be related to those of a
state-dependent single-spin Hamiltonian
\begin{equation} \label{Hcond}
 \hat{H}_{\rm condensate}= -2\alpha N \hat{J}^2+ 3 \alpha N
 \langle\hat{\vv J}\rangle \cdot \hat{\vv J}\,.
\end{equation}
It generates the non-linear equation
\begin{equation} \label{NonLin}
 i\hbar \frac{{\rm d}\chi}{{\rm d}t} = \hat{H}\chi = -2\alpha N
 \hat{J}^2\chi+ 3 \alpha N  \langle\chi|\hat{\vv J}|\chi\rangle
 \cdot \hat{\vv J} \chi\,.
\end{equation}
Unlike in the minimal minisuperspace model, the condensate dynamics is
non-trivial.

\subsection{Two interacting minisuperspace models}

The inhomogeneous dynamics can be probed more faithfully by using more than
one triple of degrees of freedom, but still of small number for a
minisuperspace model. Starting with the unreduced Hamiltonian, we split the
spins into different subsets by introducing the following notation:
\begin{eqnarray}
 \hat{\vv J}_{{\rm h}1,j}&:=&\hat{\vv J}_{4j-3}\\
\hat{\vv J}_{{\rm v}1,j}&:=& \hat{\vv J}_{4j-2}\\
\hat{\vv J}_{{\rm h}2,j}&:=&\hat{\vv J}_{4j-1}\\
\hat{\vv J}_{{\rm v}2,j}&:=&\hat{\vv J}_{4j} \, .
\label{split_spins}
\end{eqnarray}
No analogous version of such a reduction has been used in quantum cosmology
yet, but we will see that, in the present model, it can improve the simpler
reduction given by $\hat{H}^{(1)}_{\rm mini}$.

\begin{figure}[htbp]
\begin{center}
 \includegraphics[scale=1]{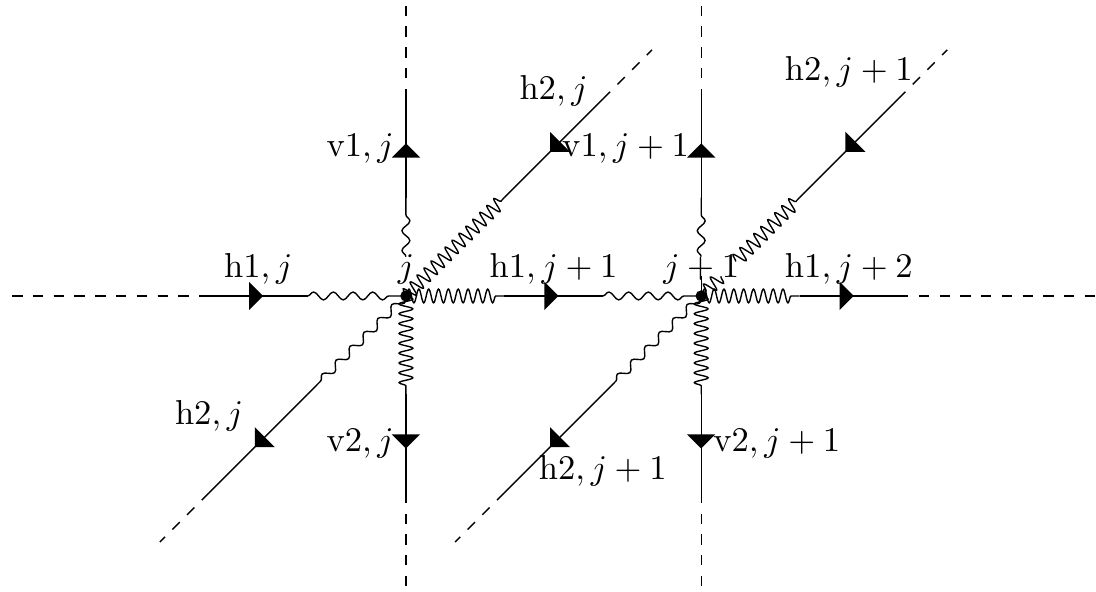}
\caption{A re-arranged representation of the discrete spin model. As before,
  lines with a given orientation stand for the spin operators. There are now
  five different spins meeting at a given vertex $j$, among which only the two
  triplets $(\hat{\vv J}_{{\rm h}1,j}, \hat{\vv J}_{{\rm v}1,j}, \hat{\vv
    J}_{{\rm h}2,j})$ and $(\hat{\vv J}_{{\rm h}2,j},\hat{\vv J}_{{\rm
      v}2,j},\hat{\vv J}_{{\rm h}1,j+1})$ are interacting as indicated by wavy
  lines with different wave lengths. The spin $\hat{\vv J}_{{\rm h}2,j}$
  appears in both triplets, and is therefore doubled in the diagrammatic
  visualization. Also the constraints relate only the spins that occur
  together in a triplet.  \label{f:Min1}}
\end{center}
\end{figure}

Assuming even $N$, we can pair up neighboring vertices ($i=2j-1$ and $i=2j$ in
(\ref{H})), and obtain the Hamiltonian
\begin{eqnarray}
 \hat{H}&=&\alpha\sum_{j=1}^{N/2} \left(-\hat{\vv J}_{4j-3}\cdot\hat{\vv J}_{4j-2}+
   \hat{\vv J}_{4j-3}\cdot\hat{\vv J}_{4j-1}+ \hat{\vv J}_{4j-2}\cdot\hat{\vv
     J}_{4j-1}\right.\\ 
&&\left.   -\hat{\vv J}_{4j-1}\cdot\hat{\vv J}_{4j}+ 
   \hat{\vv J}_{4j-1}\cdot\hat{\vv J}_{4j+1}+ \hat{\vv J}_{4j}\cdot\hat{\vv
     J}_{4j+1}\right)\\ 
&=& \alpha\sum_{j=1}^{N/2} \left(-\hat{\vv J}_{{\rm h}1,j}\cdot\hat{\vv
    J}_{{\rm v}1,j}+ 
   \hat{\vv J}_{{\rm h}1,j}\cdot\hat{\vv J}_{{\rm h}2,j}+ \hat{\vv J}_{{\rm
       v}1,j}\cdot\hat{\vv J}_{{\rm h}2,j} \right.\\ 
&&\left. -\hat{\vv J}_{{\rm h}2,j}\cdot\hat{\vv J}_{{\rm v}2,j}+
   \hat{\vv J}_{{\rm h}2,j}\cdot\hat{\vv J}_{{\rm h}1,j+1}+ \hat{\vv J}_{{\rm
       v}2,j}\cdot\hat{\vv J}_{{\rm h}1,j+1}\right)\,. 
\end{eqnarray}
(We refer to $\hat{\vv J}_{2N+1}$ as a single spin $\hat{\vv J}_{{\rm
    h}1,N/2+1}$ without vertical or a second horizontal spins for $j=N/2+1$.)
The new configurations and their interactions can be illustrated as in
Fig.~\ref{f:Min1}.

We have two constraints for each value of $j$:
\begin{eqnarray}
 \hat{C}_{1,j} &=& -\hat{\vv J}_{4j-3}+ \hat{\vv J}_{4j-2}+ \hat{\vv J}_{4j-1} =
 -\hat{\vv J}_{{\rm h}1,j}+ \hat{\vv J}_{{\rm v}1,j}+ \hat{\vv J}_{{\rm h}2,j}\\
 \hat{C}_{2,j} &=& -\hat{\vv J}_{4j-1}+ \hat{\vv J}_{4j}+ \hat{\vv J}_{4j+1} =
 -\hat{\vv J}_{{\rm h}2,j}+ \hat{\vv J}_{{\rm v}2,j}+ \hat{\vv J}_{{\rm h}1,j+1}\,.
\end{eqnarray}
The reduced Hamiltonian is
\begin{eqnarray}
\hat{H}_{{\rm red}} &=& \alpha \sum_{j=1}^{N/2} \left(-\hat{J}_{{\rm h}1,j}^2
  -2 \hat{J}_{{\rm h}2,j}^2- \hat{J}_{{\rm h}1,j+1}^2+ 3 \, \hat{\vv J}_{{\rm
      h}1,j}\cdot \hat{\vv J}_{{\rm h}2,j}+ 3 \, \hat{\vv J}_{{\rm h}2,j}\cdot
  \hat{\vv J}_{{\rm h}1,j+1}\right)\,.
\end{eqnarray}
Setting $\hat{\vv J}_{{\rm h}1,j}=\hat{\vv J}_{{\rm
    h}1}$ and $\hat{\vv J}_{{\rm h}2,j}=\hat{\vv
  J}_{{\rm h}2}$ for all $j$ (but $\hat{\vv J}_{{\rm h}1}\not=\hat{\vv
  J}_{{\rm h}2}$), we obtain a new minisuperspace Hamiltonian
\begin{equation}
 \hat{H}_{\rm mini}^{(2)} = \alpha N (-\hat{J}_{{\rm h}1}^2- \hat{J}_{{\rm
     h}2}^2+ 3 \, \hat{\vv J}_{{\rm h}1}\cdot\hat{\vv J}_{{\rm h}2})
\end{equation}
with non-trivial dynamics. We note that this minisuperspace model is
closely related to the mean-field model introduced for finite
Heisenberg spin chains in \cite{FiniteHeisenberg}. The main difference
is that our minisuperspace treatment identifies spin degrees of
freedom on alternating links, while the mean-field treatment couples
these spins. This relation, which we do not pursue further in this
paper (except for one conclusion drawn in Sec.~\ref{s:QC}), could be
useful in an extension of discrete minisuperspace models to controlled
mean-field theories.

Alternatively, we can split up the range of $N$ vertices into two disjoint
averaging regions. 
Unlike $\hat{H}^{(2)}_{\rm mini}$, such a reduction can be interpreted as an
analog of reductions proposed in quantum cosmology. In an attempt to include
degrees of freedom relevant for the evolution of inhomogeneous perturbations
on an isotropic background cosmology, \cite{ScalarHolEv} proposed that
independent spatial regions can be pasted together to allow for more general
degrees of freedom. (Our model here has only two such regions, but we will
comment on effects of subdivisions in our discussion of stability in
Sections~\ref{s:Small} and \ref{s:QC}.) Such a quantum cosmology is a version
of the classical separate-universe approximation of
\cite{SeparateUniverse,SeparateUniverseII,SeparateUniverseIII}.

In the reduced theory, we now assume $N$ odd, such that we
have an even number $N+1$ of horizontal spins. We can group them in two sets,
one for the spins around vertices $i=1$ to $i=(N-1)/2$ and one for vertices
from $i=(N+3)/2$ to $i=N$. (The central vertex $i_{\rm c}=(N+1)/2$ is not
included in this counting. The spin $\hat{\vv J}_{2i_{\rm c}-1}=\hat{\vv J}_N$
to its left is contained in the first set, while $\hat{\vv J}_{2i_{\rm
    c}+1}=\hat{\vv J}_{N+2}$ to its right is contained in the second set.)
Calling the first $(N+1)/2$ horizontal spins $\hat{\vv J}_{{\rm h}1}:=\hat{\vv
  J}_{1}=\hat{\vv J}_3\cdots=\hat{\vv J}_{N}$ and the last $(N+1)/2$ spins
$\hat{\vv J}_{{\rm h}2}:= \hat{\vv J}_{N+2}=\hat{\vv J}_{N+4}=\cdots= \hat{\vv
  J}_{2N+1}$, we obtain, starting from the reduced Hamiltonian, the
minisuperspace Hamiltonian
\begin{equation} \label{Hmini3}
 \hat{H}_{\rm mini}^{(3)} = \alpha \frac{N-3}{2} (\hat{J}_{{\rm h}1}^2+\hat{J}_{{\rm
     h}2}^2) +3\alpha \hat{\vv J}_{{\rm h}1}\cdot\hat{\vv J}_{{\rm h}2}\,.
\end{equation}
(There are $(N-1)/2$ non-interacting contributions of $\hat{J}_{{\rm h}1}^2$
and $\hat{J}_{{\rm h}2}^2$ from spins in the interior of the two averaging
regions, as well as one contribution of $-\hat{J}_{{\rm h}1}^2+3\hat{\vv
  J}_{{\rm h}1}\cdot\hat{\vv J}_{{\rm h}2} -\hat{J}_{{\rm h}2}^2$ with
interactions at the border between the regions, located at the central vertex
$i_{\rm c}=(N+1)/2$.)

The two Hamiltonians are rather different from each other. They are both of
the form 
\begin{equation} \label{Hgeneric}
\hat{H}_{\beta\gamma}:= \beta(\hat{J}_{{\rm h}1}^2+\hat{J}_{{\rm h}2}^2)+\gamma
\hat{\vv J}_{{\rm h}1}\cdot\hat{\vv J}_{{\rm h}2}\,,
\end{equation}
but while ${\rm sgn}\gamma={\rm sgn}\alpha$ in both cases, we have ${\rm
  sgn}\beta=-{\rm sgn}\alpha$ for $\hat{H}^{(2)}$ and ${\rm sgn}\beta={\rm
  sgn}\alpha$ for $\hat{H}^{(3)}$ (with $N>3$). We should therefore expect
different ground states or effective potentials in the two cases. In
particular, if $\langle\hat{J}_{{\rm h}1/2}^2\rangle$ is considered a free
variable, we can minimize the energy of $\langle\hat{H}^{(3)}\rangle$ by zero
spins, while the energy range of $\langle\hat{H}^{(2)}\rangle$ is unbounded
from below. If $\langle\hat{J}_{{\rm h}1/2}^2\rangle$ is fixed, however, only
the interaction term matters for ground states or effective potentials, for
which the two models provide the same sign. The ground-state properties are
then similar to those of the Heisenberg spin chain related to the reduced
Hamiltonian (\ref{Hreduced}): parallel alignment of next-neighbor spins
(ferromagnetic) if $\gamma<0$ and antiparallel alignment (antiferromagnetic)
if $\gamma>0$.

We assume that the individual spin states have the same eigenvalue of
$\hat{J}_{{\rm h}1}^2$ and $\hat{J}_{{\rm h}2}^2$, given by $s(s+1)\hbar^2$
with some half-integer $s$. We have two cases according to ${\rm sgn}\gamma$:
For $\gamma>0$, the interaction term is minimized by antiparallel spins
$\hat{\vv J}_{{\rm h}1/2}$.  The two spins then form a combined spin
eigenstate $|0,0\rangle$ of the total spin $\hat{\vv J}:=\hat{\vv J}_{{\rm
    h}1}+\hat{\vv J}_{{\rm h}2}$ in which
\begin{equation}
 \langle\hat{\vv J}_{{\rm
    h}1}\cdot\hat{\vv J}_{{\rm h}2}\rangle= \frac{1}{2}
\langle\hat{J}^2-\hat{J}_{{\rm h}1}^2-\hat{J}_{{\rm h}2}^2\rangle=
-s(s+1)\hbar^2\,,
\end{equation}
and the energy eigenvalue is given by
\begin{equation} \label{Epos}
 E_{\gamma>0} = (2\beta-\gamma) s(s+1)\hbar^2\,.
\end{equation}
For $\gamma<0$, the parallel configuration minimizes the interaction term, for
which we have a whole multiplet of different sates with total spin $2s$.  In
any such state, $\langle\hat{\vv J}_{{\rm h}1}\cdot \hat{\vv J}_{{\rm
    h}2}\rangle=s^2\hbar^2$, and the energy eigenvalue is
\begin{equation} \label{Eneg}
 E_{\gamma<0} = (2\beta s(s+1)+ \gamma s^2) \hbar^2\,.
\end{equation}

The two different types of ground states, with antiparallel spins for
$\gamma>0$ and parallel spins for $\gamma<0$, have interesting implications
for the reliability of the two different interacting mini\-superspace
models. If $\gamma<0$, the two models predict the same ground-state
configuration with all fundamental spins aligned. There is only a quantitative
difference between the models in the predicted ground-state energy. For
$\gamma>0$, however, the antiparallel alignment of $\hat{\vv J}_{{\rm h}1}$
and $\hat{\vv J}_{{\rm h}2}$ corresponds to very different fundamental
configurations. With $\hat{H}_{\rm mini}^{(2)}$, the two antiparallel spins
are alternating along the full spin chain, which agrees with the ground state
of the discrete theory. With $\hat{H}_{\rm mini}^{(3)}$, however, we have two
averaging regions with equal spins in each region, but antiparallel alignment
between the two regions. Knowing the fundamental configuration, we can tell
that an energy preference of antiparallel alignment at the border between the
two regions means that the configuration should be unstable under splitting it
up further into smaller and smaller averaging regions with antiparallel
alignment at all borders. With complete splitting, each link being an
averaging region of its own, a configuration as with $\hat{H}_{\rm
  mini}^{(2)}$ or in the fundamental theory is obtained, but one would have
left the minisuperspace stage. The second minisuperspace model with
Hamiltonian $\hat{H}_{\rm mini}^{(2)}$, on the other hand, realizes the
correct ground state within a minisuperspace model, and without the need for
further refinement. (There is, however, a difference between the ground state
of $\hat{H}_{\rm mini}^{(2)}$ and the corresponding discrete theory. The
former has an even number of spins, but it is derived from a spin chain with
an odd number of spins. These two cases are known to have different behaviors
\cite{BonnerFisher,SpinChainEvenOdd,EvenOddHeisenberg}.)

We conclude that the way degrees of freedom are included in a minisuperspace
model can have significant implications for how well fundamental properties
are modelled, which however can be evaluated only if one knows a great deal
about the fundamental theory. Transferring this lesson to quantum gravity
suggests that caution toward minisuperspace results would be
advisable. However, there is a difference between stability as discussed so
far, where it is implicitly assumed that the spin chain can exchange energy
with an environment and settle down to its ground state, and quantum-cosmology
models, where there is no environment outside of the system. We will return to
this question in Secs.~\ref{s:EffDisc} and \ref{s:QC}.

\subsection{Effective equations and potentials}

We continue to analyze the dynamics by means of effective equations in
canonical form. Following \cite{EffAc,Karpacz,Counting}, we assign infinitely
many numbers to a set of quantum spin degrees of freedom with operator
$\hat{\vv J}_i$. In a given state, these numbers correspond to the expectation
values $\langle\hat{J}^a_i\rangle$ of spin components and the moments
\begin{equation}
 \Delta(J^{a_1}_{i_1}\cdots J^{a_n}_{i_n}) := \langle
 (\hat{J}^{a_1}_{i_1}-\langle\hat{J}^{a_1}_{i_1}\rangle) \cdots
 (\hat{J}^{a_n}_{i_n}-\langle\hat{J}^{a_n}_{i_n}\rangle)\rangle_{\rm symm}
\end{equation}
in totally symmetric ordering. (For uniform notation of all moments, we write
fluctuations as $\Delta(J^aJ^a)=(\Delta J^a)^2$.) In a semiclassical
expansion, which we will perform in most of our derivations, it is sufficient
to include only moments up to second order, $n=2$. We then have a
finite-dimensional system. For a semiclassical state, defined as a state with
moments of the order $\Delta(J^{a_1}_{i_1}\cdots J^{a_n}_{i_n})=
\mathcal{O}(\hbar^{(a_1+\cdots+a_n)/2})$, terms of order $\hbar$ are included in a
truncation up to second order in moments.

The quantum degrees of freedom form a phase space with Poisson bracket
derived from the commutator via
\begin{equation}
 \{\langle\hat{A}\rangle,\langle\hat{B}\rangle\} :=
 \frac{\langle[\hat{A},\hat{B}]\rangle}{i\hbar}\,.
\end{equation}
For instance,
\begin{eqnarray}
 \{\langle\hat{J}_i^a\rangle,\langle\hat{J}_j^b\rangle\} &=& \epsilon^{abc}
 \langle\hat{J}_i^c\rangle \delta_{ij}\nonumber\\
 \{\Delta(J_i^aJ_j^b),\langle\hat{J}_k^c\rangle\} &=&
 \delta_{ik}\epsilon^{acd} \Delta(J_i^dJ_j^b)+ \delta_{jk}
 \epsilon^{bcd}\Delta(J_i^aJ_j^d)\label{PB}\\
 \{\Delta(J_i^aJ_j^b),\Delta(J_k^cJ_l^d)\} &=& \delta_{ik}\epsilon^{ace}
 \langle\hat{J}_i^e\rangle  \Delta(J_j^bJ_l^d)+ \delta_{il}
 \epsilon^{ade}\langle\hat{J}_i^e\rangle \Delta(J_j^bJ_k^c) + \delta_{jk}
 \epsilon^{bce} \langle\hat{J}_j^e\rangle \Delta(J_i^aJ_l^d) \nn \\
 &+& \delta_{jl} \epsilon^{bde} \langle\hat{J}_j^e\rangle \Delta(J_i^aJ_k^c)+ \mathcal{O}(\Delta^3)\,.
\nonumber
\end{eqnarray}
The first two lines are exact, while the third line is valid to second order
in moments.

The Poisson bracket allows us to compute equations of motion. For instance, in
the minimal minisuperspace model we have the effective Hamiltonian
\begin{equation} \label{Hmini1eff}
 H_{\rm eff}^{(1)} = \alpha N \left(\langle\hat{J}^x\rangle^2+
 \langle\hat{J}^y\rangle^2+\langle\hat{J}^z\rangle^2+
 \Delta(J^xJ^x)+\Delta(J^yJ^y)+ \Delta(J^zJ^z)\right)\,.
\end{equation}
All expectation values and moments then have zero time derivatives 
\begin{equation}
 \frac{{\rm d} O}{{\rm d}t} = \{O,H_{\rm eff}^{(1)}\}\,,
\end{equation}
as is to be expected from the non-interacting nature of the system.

\subsubsection{Minimal minisuperspace model}

The moments in (\ref{Hmini1eff}) provide a quantum correction to the classical
Hamiltonian. For the ground state, we can derive their minimal values by
saturating uncertainty relations
\begin{equation}
 \Delta(J^xJ^x)\Delta(J^yJ^y)\geq \frac{\hbar^2}{4}
 |\langle\hat{J}^z\rangle|^2
\end{equation}
and cyclic permutations. In order to evaluate these equations, we may assume
that the expectation value $\langle\hat{\vv J}\rangle$ points in the
$z$-direction. Therefore, only $\langle\hat{J}^z\rangle$ is non-zero among the
spin components. The saturated uncertainty relations then require that
$\Delta(J^zJ^z)=0$ while 
\begin{equation} \label{Deltaxx}
 \Delta(J^xJ^x)\Delta(J^yJ^y)=\frac{1}{4}\hbar^2 |\langle\hat{J}^z\rangle|^2\,.
\end{equation}
By symmetry, $\Delta(J^xJ^x)=\Delta(J^yJ^y)$. The quantum correction to the
classical Hamiltonian, or the effective potential, is then
\begin{equation} \label{Veff1}
  V_{\rm eff}^{(1)} = \alpha N(\Delta(J^xJ^x)+\Delta(J^yJ^y)+\Delta(J^zJ^z))=
  \alpha N \hbar |\langle\hat{J}^z\rangle|=
  \alpha N \hbar\sqrt{\delta_{ab}\langle\hat{J}^a\rangle
    \langle\hat{J}^b\rangle}\,. 
\end{equation}
Although the effective potential is non-zero, it does not imply interactions
but rather provides a zero-point energy because it only depends on a constant
of motion. The effective Hamiltonian is given by
\begin{equation} \label{Heff1}
 H_{\rm eff}^{(1)} = \alpha N \langle\hat{J}^z\rangle^2+ V_{\rm eff}^{(1)} =
 \alpha N \left( \langle\hat{J}^z\rangle^2+ \hbar 
   |\langle\hat{J}^z\rangle|\right)= \alpha N s(s+1)\hbar^2
\end{equation}
if $\langle\hat{J}^z\rangle =\pm s \hbar$ for a spin pointing in the
$z$-direction according to our assumptions. This value of the effective
Hamiltonian is in agreement with the operator result, where the well-known
eigenvalues of $\hat{J}^2$ are $s(s+1)\hbar^2$.

The same effective potential can be obtained from a canonical version of the
spin system. Again assuming that the spin vector points in the $z$-direction,
we introduce canonical coordinates
\begin{eqnarray}
 q:= \frac{\delta\langle\hat{J}^x\rangle}{\sqrt{\langle\hat{J}^z\rangle}}=
 \frac{\delta\langle\hat{J}^x\rangle}{\sqrt[4]{\delta_{ab}
     \langle\hat{J}^a\rangle\langle\hat{J}^b\rangle-
   (\delta\langle\hat{J}^x\rangle)^2- 
     (\delta\langle\hat{J}^y\rangle)^2}}  \nn \\
  p:=
 \frac{\delta\langle\hat{J}^y\rangle}{\sqrt{\langle\hat{J}^z\rangle}}=
 \frac{\delta\langle\hat{J}^y\rangle}{\sqrt[4]{\delta_{ab}
     \langle\hat{J}^a\rangle\langle\hat{J}^b\rangle-
 (\delta\langle\hat{J}^x\rangle)^2-
     (\delta\langle\hat{J}^y\rangle)^2}} 
\end{eqnarray}
for small values $\delta\langle\hat{J}^x\rangle$ and
$\delta\langle\hat{J}^y\rangle$ much less than
$\langle\hat{J}^z\rangle$: We have 
\begin{equation}
 \{q,p\} = \frac{\{\delta \langle\hat{J}^x\rangle,\delta
   \langle\hat{J}^y\rangle\}}{\langle\hat{J}^z\rangle}+ 
   \frac{\delta \langle\hat{J}^x\rangle\{\delta
     \langle\hat{J}^y\rangle,\langle\hat{J}^z\rangle\}-
     \delta\langle\hat{J}^y\rangle \{\delta\langle\hat{J}^x\rangle,
     \langle\hat{J}^z\rangle\}}{2\langle\hat{J}^z\rangle^2}=
   1+\mathcal{O}((\delta\langle\hat{J}^{x/y}\rangle/\langle\hat{J}^z\rangle)^2)\,.
\end{equation}
These relations imply the quadratic equation
\begin{equation} \label{Quad}
 \left((\delta\langle\hat{J}^x\rangle)^2+
 (\delta\langle\hat{J}^y\rangle)^2\right)^2+ (q^2+p^2)^2 
 \left((\delta\langle\hat{J}^x\rangle)^2+
(\delta\langle\hat{J}^y\rangle)^2\right)-
 (q^2+p^2)^2 \delta_{ab} 
 \langle\hat{J}^a\rangle \langle\hat{J}^b\rangle=0\,.
\end{equation}
For $\delta_{ab} \langle\hat{J}^a\rangle \langle\hat{J}^b\rangle$ much greater
than $(\delta\langle\hat{J}^x\rangle)^2+(\delta\langle\hat{J}^y\rangle)^2$, we
can solve (\ref{Quad}) for
$(\delta\langle\hat{J}^x\rangle)^2+(\delta\langle\hat{J}^y\rangle)^2$ by
\begin{eqnarray}
 (\delta\langle\hat{J}^x\rangle)^2+(\delta\langle\hat{J}^y\rangle)^2 &=&
 -\frac{1}{2}  (q^2+p^2)^2\pm \sqrt{\delta_{ab} \langle\hat{J}^a\rangle
   \langle\hat{J}^b\rangle} (q^2+p^2) \sqrt{1+\frac{(q^2+p^2)^2}{4\delta_{ab}
     \langle\hat{J}^a\rangle \langle\hat{J}^b\rangle}} \nn \\
     & \approx &
 \sqrt{\delta_{ab} \langle\hat{J}^a\rangle \langle\hat{J}^b\rangle} (q^2+p^2)-
 \frac{1}{2} (q^2+p^2)^2 +\cdots\,.
\end{eqnarray}
(Only the plus sign gives a positive solution.)  The first term is a
harmonic-oscillator Hamiltonian with $m^{-1}=2\sqrt{\delta_{ab}
  \langle\hat{J}^a\rangle \langle\hat{J}^b\rangle}=\omega$, which has
zero-point energy $\frac{1}{2}\hbar\omega= \hbar\sqrt{\delta_{ab}
  \langle\hat{J}^a\rangle \langle\hat{J}^b\rangle}$ in agreement with the
effective potential (\ref{Veff1}).

\subsubsection{Interacting minisuperspace models}

For the two interacting models, we need the covariances $\Delta(J_{{\rm
    h}1}^aJ_{{\rm h}2}^a)$ for $a=x,y,z$ (no sum over $a$). In addition to the
fluctuations as in the minimal minisuperspace model with a Hamiltonian of the
type (\ref{Hgeneric}), we then have the term
\begin{equation} \label{Vint}
  \Delta V_{\rm interaction}=\gamma (\Delta(J_{{\rm h}1}^xJ_{{\rm h}2}^x)+
  \Delta(J_{{\rm h}1}^yJ_{{\rm 
      h}2}^y)+ \Delta(J_{{\rm h}1}^zJ_{{\rm h}2}^z))
\end{equation}
in the effective potential.

Covariances can often be ignored in the context of uncertainty relations, but
they do contribute in the complete form
\begin{equation}
 \Delta(A^2)\Delta(B^2)-\Delta(AB)^2\geq \frac{1}{4}
 |\langle[\hat{A},\hat{B}]\rangle|^2\,.
\end{equation}
Unlike fluctuations, which obtain a lower bound for non-commuting operators,
covariances are subject to an upper bound depending on the fluctuations. For
two commuting spin components as in (\ref{Vint}), the uncertainty relation is
\begin{equation}
 \Delta(J_{{\rm h}1}^xJ_{{\rm h}1}^x)\Delta(J_{{\rm h}2}^xJ_{{\rm h}2}^x) \geq
 \Delta(J_{{\rm h}1}^xJ_{{\rm h}2}^x)^2\,.
\end{equation}

If $\gamma>0$ in (\ref{Vint}), we can minimize the effective interaction
potential by choosing the value
\begin{equation} \label{Delta12}
\Delta(J_{{\rm h}1}^xJ_{{\rm h}2}^x)=- \sqrt{\Delta(J_{{\rm
      h}1}^xJ_{{\rm h}1}^x)\Delta(J_{{\rm h}2}^xJ_{{\rm h}2}^x)}=
-\frac{1}{2}\hbar |\langle\hat{J}^z\rangle|\,.
\end{equation}
The last equality holds if we also minimize the uncertainty relations for the
two individual spins and use the antiparallel alignment for spin expectation
values, without loss of generality assumed to point in the $z$-direction. The
same value is obtained for $\Delta(J_{{\rm h}1}^yJ_{{\rm h}2}^y)$, while
$\Delta(J_{{\rm h}1}^zJ_{{\rm h}2}^z)=0$ because the $z$-components have zero
fluctuations with our choice of directions. Combining all terms in the
effective Hamiltonian (\ref{Hgeneric}) then yields
\begin{equation}
 H_{{\rm eff},\gamma>0} = 2\beta s(s+1)\hbar^2 - \gamma(s^2\hbar^2+ s\hbar^2)=
 s(s+1)(2\beta-\gamma) \hbar^2\,.
\end{equation}
The first term is twice the non-interacting contribution from a single spin
with the same form as in (\ref{Heff1}), the second term, $-\gamma s^2\hbar^2$,
is $\gamma \langle\hat{J}_{{\rm h}1}^z\rangle \langle\hat{J}_{{\rm
    h}2}^z\rangle$ for antiparallel alignment in the $z$-direction, and the
last term, $-\gamma s\hbar^2$, adds up the two non-zero covariances in
(\ref{Vint}). This result agrees with the ground-state energy
(\ref{Epos}). For $\gamma<0$, one can see in the same way that the operator
result (\ref{Eneg}) is reproduced if the covariances vanish and parallel
alignment is used in $\gamma \langle\hat{J}_{{\rm h}1}^z\rangle
\langle\hat{J}_{{\rm h}2}^z\rangle=\gamma s^2\hbar^2$.

\subsubsection{Condensate model}

The condensate model has provided non-trivial dynamics for a single
spin. However, some of the equations for expectation values and moments are
trivial. We can derive effective equations from the state-dependent
Hamiltonian (\ref{Hcond}) if we assign to it the effective Hamiltonian
\begin{equation}
 H_{\rm eff,\: cond}= -2\alpha N \delta_{ab} \left(\langle\hat{J}^a\rangle
 \langle\hat{J}^b\rangle+ \Delta(J^aJ^b)\right)+ 3\alpha N \delta_{ab} j^a
\langle\hat{J}^b\rangle
\end{equation}
where the vector $j^a$ is treated in the following way: It is a constant for
purposes of computing Poisson brackets, for instance in equations of
motion. After the Poisson brackets have been derived, one sets
$j^a=\langle\hat{J}^a\rangle$. In this way, the correct equations follow for
expectation values taken in a state that evolves according to the non-linear
equation (\ref{NonLin}). 

Following this procedure, we obtain the equation
\begin{equation}
 \frac{{\rm d}\langle\hat{J}^c\rangle}{{\rm d}t} = 3\alpha N \epsilon^{abc}
 \langle\hat{J}^a\rangle \langle\hat{J}^b\rangle=0\,.
\end{equation}
For second-order moments, we have the contribution
\begin{equation}
 \frac{{\rm d}\langle\hat{J}^c\hat{J}^d\rangle}{{\rm d}t} = 3\alpha N
 \langle\hat{J}^a\rangle \left\langle\epsilon^{abc} \hat{J}^b\hat{J}^d+
   \epsilon^{abd} \hat{J}^c\hat{J}^b\right\rangle\,.
\end{equation}
Since covariances couple to expectation values and fluctuations, non-zero
correlations can build up during evolution even if they vanish in an initial
state.

\section{Continuum theories}

One question about minisuperspace models derived from classical continuum
theories, addressed in \cite{MiniSup}, is about the coordinate volume $V_0$
that characterizes the size of an averaging region. (See \cite{ROPP} for a
review of minisuperspace models and quantum cosmology.) For the simplest
minisuperspace models of general relativity, space is flat and infinite, and
the canonical form $\int {\rm d}^3x \dot{\phi}p_{\phi}$ of any local degree of
freedom $\phi(x)$ with momentum $p_{\phi}(x)$ is infinite for homogeneous
configurations. In order to obtain a well-defined canonical structure, one can
choose a finite region ${\cal V}$ of coordinate size $V_0=\int_{\cal V}{\rm
  d}^3x$ and restrict integrations of the canonical form and the Lagrangian to
this region.

If only homogeneous configurations are considered, the size and position of
this region should not matter. The restricted canonical form is
\begin{equation}
 \int_{\cal V}{\rm d}^3x\dot{\phi}(x)p_{\phi}(x)= V_0\dot{\bar{\phi}}\bar{p}_{\phi}
\end{equation}
for homogeneous configurations $\phi(x)=\bar{\phi}$ and
$p_{\phi}(x)=\bar{p}_{\phi}$. The momentum of $\bar{\phi}$ is therefore not
equal to $\bar{p}_{\phi}$, but to $\bar{p}=V_0\bar{p}_{\phi}$. The standard
Hamiltonian of a scalar field, just like other Hamiltonians for instance of
gravitational degrees of freedom, then depends on $V_0$ when it is restricted
to canonically conjugate minisuperspace configurations. For a scalar field, we
have
\begin{equation}
 \int_{\cal V}{\rm d}^3x \left(\frac{1}{2} \dot{\bar{\phi}}^2
   +W(\bar{\phi})\right) =
 \frac{V_0}{2}\bar{p}_{\phi}^2+V_0W(\bar{\phi})= \frac{1}{2}
 \frac{\bar{p}^2}{V_0}+ V_0W(\bar{\phi})
\end{equation}
and the dependence on $V_0$ is not just by a multiplicative factor. While the
classical theory does not depend on the choice of $V_0$, the quantum theory
does, for instance via $V_0$-dependent effective potentials.

In \cite{MiniSup}, it was shown that the $V_0$-dependent semiclassical
corrections in minisuperspace effective potentials of a scalar-field theory
are related to infrared contributions to field-theory effective
potentials. The choice of $V_0$ then has physical relevance in the number of
infrared modes included in the minisuperspace model, but it is a property of
the minisuperspace truncation rather than of physical interactions. It is
therefore difficult to justify interpretations of minisuperspace effects in
quantitative terms.

In order to test this question in our spin system, we now derive a continuum
theory from which our minimal minisuperspace model can be obtained by using
homogeneous configurations. Generalizations corresponding to the interacting
minisuperspace models will also be considered. As we will see, for this kind
of spin systems the nature of averaging regions is less problematic than for
canonical field theories. There may therefore be an advantage in deriving
minisuperspace models directly from discrete quantum theories.

\subsection{Continuum models}

Starting with the unreduced theory, we introduce two continuum fields
$\hat{\vv J}_{\rm h}(x)$ and $\hat{\vv J}_{\rm v}(x)$ where $x$ runs through the
entire length of our graph. In terms of the measure provided by the choice of
$x$, we introduce the coordinate distance between two vertices $i$ at $x=x_i$
and $i+1$ at $x=x_{i+1}$ by $\delta=\int_{x_i}^{x_{i+1}}{\rm d}x$, assumed to
be independent of $i$. The total length of the graph is
$L_0=\int_{x_1}^{x_{N+1}}{\rm d}x = N \delta$.  For integer values of $x=x_i$,
we identify $\hat{\vv J}_{\rm v}(x_i)=\hat{\vv J}_{{\rm v},i}= \hat{\vv J}_{2i}$,
$\hat{\vv J}_{{\rm h}1,i}= \hat{\vv J}_{2i-1}$ and $\hat{\vv J}_{{\rm h}2,i}=
\hat{\vv J}_{2i+1}$.

\subsubsection{Minimal model}

For the minimal model, we do not treat $\hat{\vv J}_{{\rm h}1,i}$ and
$\hat{\vv J}_{{\rm h}2,i}$ as independent fields but rather view them as one
horizontal field evaluated at different positions $x_i\pm\delta/2$: We set
$\hat{\vv J}_{\rm h}(x_i-\delta/2)=\hat{\vv J}_{{\rm h}1,i}$ and $\hat{\vv J}_{\rm
  h}(x_i+\delta/2)=\hat{\vv J}_{{\rm h}2,i}$. The leading terms in a continuum
limit of the Hamiltonian and the constraint can then be obtained by an
expansion in $\delta$ up to second order. Derivatives by $x$ appear in the
process, and will be denoted by a prime.

The constraints are
\begin{eqnarray}
 \hat{C}^a(x_i) &=& -\hat{J}_{\rm h}^a(x_i-\delta/2)+ \hat{J}_{\rm v}^a(x_i)+
 \hat{J}_{\rm h}^a(x_i+\delta/2)\\
 &=& \hat{J}_{\rm v}^a(x_i)+ \delta \hat{J}_{\rm h}^a(x_i)'\,.
\end{eqnarray}
The Hamiltonian is
\begin{eqnarray}
 \hat{H} &=& \alpha \delta_{ab}\sum_{i=1}^N \left(-\hat{J}_{\rm
     h}^a(x_i-\delta/2) \hat{J}_{\rm v}^b(x_i)+ \hat{J}_{\rm
     h}^a(x_i-\delta/2) \hat{J}_{\rm h}^b(x_i+\delta/2)+ \hat{J}_{\rm
     v}^a(x_i) \hat{J}_{\rm h}^b(x_i+\delta/2)\right)\nn \\
 &=& \alpha \sum_{i=1}^N \left(\hat{J}_{\rm h}(x_i)^2+ \delta \hat{\vv J}_{\rm
     v}(x_i)\cdot \hat{\vv J}_{\rm h}(x_i)'+ \frac{1}{4}\delta^2 (\hat{\vv J}_{\rm
     h}(x_i)\cdot \hat{\vv J}_{\rm h}(x_i)''- (\hat{J}_{\rm h}(x_i)')^2)\right)\,.
\end{eqnarray}
Solving the constraint, the Hamiltonian becomes
\begin{equation}
 \hat{H} = \alpha \sum_{i=1}^N \left(\hat{J}_{\rm h}(x_i)^2+
   \frac{1}{4}\delta^2 (\hat{\vv J}_{\rm h}(x_i)\cdot \hat{\vv J}_{\rm h}(x_i)''-
   5(\hat{J}_{\rm h}(x_i)')^2)\right)
\end{equation}
and gives rise to the continuum limit
\begin{eqnarray}
 \hat{H}_{\rm cont}^{(1)} &=& \frac{\alpha}{\delta} \int{\rm d}x
 \left(\hat{J}_{\rm h}^2+ \frac{1}{4}\delta^2 (\hat{\vv J}_{\rm h}\cdot
   \hat{\vv J}_{\rm h}''- 5(\hat{J}_{\rm h}')^2)\right)\\
 &=& \frac{\alpha}{\delta} \int{\rm d}x \left(\hat{J}_{\rm h}^2- \frac{3}{2}
   \delta^2 (\hat{J}_{\rm h}')^2\right)+ \frac{1}{4}\alpha\delta \hat{\vv J}_{\rm
   h}\cdot \hat{\vv J}_{\rm h}'|_{\partial} \label{Hcont1}
\end{eqnarray}
with a boundary term denoted by a subscript $\partial$. The
continuum limit for the reduced Hamiltonian (\ref{Hreduced}) is the same as
(\ref{Hcont1}).

The minisuperspace model obtained from this continuum Hamiltonian is 
\begin{equation}
 H_{\rm mini}^{(1)} = \frac{\alpha}{\delta}L_0 \hat{J}_{\rm h}^2= \alpha N
 \hat{J}_{\rm h}^2\,.
\end{equation}
It is identical with the Hamiltonian in our minimal minisuperspace model. Any
reference to the averaging length $L_0$ can be expressed in terms of the
number $N$ of vertices of the fundamental discrete theory, which has physical
meaning free of truncation choices. Therefore, there are no such problems as
may be related to the appearance of $V_0$ in minisuperspace models
derived from classical continuum theories.

\subsubsection{Interacting minisuperspace models}

If we do distinguish between $\hat{\vv J}_{{\rm h}1}(x_i)$ and $\hat{\vv
  J}_{{\rm h}2}(x_i)$, we obtain the continuum Hamiltonian
\begin{equation}
 \hat{H}_{\rm cont}^{(2)} = \frac{\alpha}{\delta} \int{\rm d}x
 \left(-\hat{\vv J}_{{\rm h}1}\cdot \hat{\vv J}_{{\rm v}} + \hat{\vv J}_{{\rm
       h}1}\cdot 
   \hat{\vv J}_{{\rm h}2} + \hat{\vv J}_{\rm v} \cdot \hat{\vv J}_{{\rm h}2}\right) 
\end{equation}
with constraint 
\begin{equation}
 \hat{C}=-\hat{\vv J}_{{\rm h}1}+\hat{\vv J}_{\rm v}+ \hat{\vv J}_{{\rm h}2}=0\,.
\end{equation}
Solving the constraint gives the Hamiltonian
\begin{equation}
 \hat{H}_{\rm cont}^{(2)} = \frac{\alpha}{\delta} \int{\rm d}x
 \left(-\hat{J}_{{\rm h}1}^2 + 3 \hat{\vv J}_{{\rm h}1}\cdot
   \hat{\vv J}_{{\rm h}2} - \hat{J}_{{\rm h}2}^2\right)\,, 
\end{equation}
from which we obtain the interacting minisuperspace Hamiltonian $\hat{H}_{\rm
  mini}^{(2)}$.

For the second interacting model, we do not distinguish between $\hat{\vv
  J}_{{\rm h}1}(x_i)$ and $\hat{\vv J}_{{\rm h}2}(x_i)$, but introduce two
averaging regions of length $L_0/2$ each in which we have the constant fields
$\hat{\vv J}_1$ and $\hat{\vv J}_2$, respectively. Starting with the first
continuum theory with Hamiltonian (\ref{Hcont1}), we obtain
\begin{equation}
 \hat{H}_{\rm hom}^{(3)}=\frac{\alpha}{\delta}\frac{L_0}{2}
 \left(\hat{J}_{1}^2+\hat{J}_2^2\right)+ 
 \frac{1}{4}\alpha\delta \left(\hat{\vv J}_1-\hat{\vv J}_2\right)\cdot
 \hat{\vv J}'\,. 
\end{equation}
The last term, originating from the boundary term, needs further
discussion. We have left a $\hat{\vv J}'$ in the equation, still referring to
an inhomogeneous continuum field. It gives us the difference between the two
averaged fields $\hat{\vv J}_1$ and $\hat{\vv J}_2$. It is infinite if we have
two constant fields taking different values in the two regions, but it is
multiplied with the spacing $\delta$ which goes to zero in the continuum
limit. We can regularize this product to a finite number by defining it as
\begin{equation} \label{Reg}
 \lim_{\delta\to 0}(\delta \hat{\vv J}') =\lim_{\delta\to 0}\left( \delta
   \frac{\hat{\vv J}_2-\hat{\vv J}_1}{\delta}\right)= \hat{\vv J}_2-\hat{\vv
   J}_1\,. 
\end{equation}
The minisuperspace Hamiltonian is then equal to
\begin{eqnarray}
 \hat{H}_{\rm mini}^{(3)}&=&\frac{\alpha}{\delta}\frac{L_0}{2}
 \left(\hat{J}_{1}^2+\hat{J}_2^2\right)-
 \frac{1}{4}\alpha\left(\hat{\vv J}_1-\hat{\vv J}_2\right)^2\\
&=& \alpha (N/2-1/4) \left(\hat{J}_1^2+ \hat{J}_2^2\right) + \frac{1}{2}\alpha
\hat{\vv J}_1\cdot \hat{\vv J}_2\,.
\end{eqnarray}
Unlike what we saw for $\hat{H}_{\rm mini}^{(2)}$, this result is not
identical with the previous derivation (\ref{Hmini3}) from the discrete
theory. However, if we slightly modify our boundary regularization by
introducing an additional factor of six in the definition (\ref{Reg}), the
minisuperspace Hamiltonians are the same. The initial disagreement is a result
only of the fact that the boundary regularization is ambiguous, which is
necessary for $\hat{H}_{\rm mini}^{(3)}$ derived from the continuum theory,
but not for $\hat{H}_{\rm mini}^{(2)}$. Notice that the final agreement is
non-trivial, because changing the factor in (\ref{Reg}) modifies the
interacting as well as non-interacting terms in the resulting $\hat{H}_{\rm
  mini}^{(3)}$, which then both agree with (\ref{Hmini3}).

\section{Effective analysis of the discrete theory}
\label{s:EffDisc}

So far, we have analyzed different minisuperspace models, their ground states
and effective potentials, as well as relations with continuum
theories. Minisuperspace models of quantum gravity are used to analyze
homogeneous solutions, which necessarily have long-range correlations as seen
from the fundamental theory because distant degrees of freedom are
identified. Non-zero fluctuations of a single minisuperspace variable are
therefore the same as long-range correlations in the fundamental theory. 

Homogeneity and long-range correlations can easily be achieved in the ground
state of the fundamental theory. However, discrete ground states are not
suitable as nearly classical geometries of a discrete model of quantum
gravity, in which degrees of freedom must be excited in order to have
non-degenerate geometries. We therefore need excited states with approximate
homogeneity and long-range correlations, which poses a question very different
from just finding ground states. Long-range correlations related to
homogeneity easily build up us a system settles down to its ground state, but
it is not guaranteed that this can happen also in an excited state of an
isolated system (such as the whole universe) which does not have a drain for
surplus energy. In our discrete model, the evolution of long-range
correlations can be studied in qualitative terms.

\subsection{Effective dynamics}

For the full or reduced discrete theories, we have large systems of coupled
equations generated by the effective Hamiltonians
\begin{eqnarray}
H_{\rm eff} &=& \alpha \delta_{ab} \sum \limits_{i=1}^{N} \,
\big(-\langle\hat{J}_{2i-1}^a\rangle 
\langle\hat{J}_{2i}^b\rangle + \langle\hat{J}_{2i-1}^a\rangle
\langle\hat{J}_{2i+1}^b\rangle  + \langle\hat{J}_{2i}^a\rangle
\langle\hat{J}_{2i+1}^b\rangle   -\Delta(J_{2i-1}^aJ_{2i}^b) \nn \\
&+& \Delta(J_{2i-1}^aJ_{2i+1}^b) + \Delta(J_{2i}^aJ_{2i+1}^b)\big)  
\end{eqnarray}
and
\begin{eqnarray}
\label{H_Qred}
 H_{\rm eff, red} &=& \alpha \delta_{ab} \sum \limits_{i=1}^{N} \,
\big(-\langle\hat{J}_{2i-1}^a\rangle 
\langle\hat{J}_{2i-1}^b\rangle + 3\langle\hat{J}_{2i-1}^a\rangle
\langle\hat{J}_{2i+1}^b\rangle  - \langle\hat{J}_{2i+1}^a\rangle
\langle\hat{J}_{2i+1}^b\rangle   -\Delta(J_{2i-1}^aJ_{2i-1}^b) \nn \\
&+& 3\Delta(J_{2i-1}^aJ_{2i+1}^b) - \Delta(J_{2i+1}^aJ_{2i+1}^b)\big)\,,
\end{eqnarray}
respectively. The Poisson brackets between expectation values and moments are
as in (\ref{PB}), except that the subscript takes values in a larger set.

For the unreduced system, we have effective constraints in addition to the
effective Hamiltonian. A single constraint operator generates infinitely many
effective constraints because it restricts not only expectation values but
also the associated moments \cite{EffCons,EffConsRel}. In states annihilated
by the constraint, we have
\begin{equation}
 C_{i}^a:=\langle\hat{C}_i^a\rangle= -\langle\hat{J}_{2i-1}^a\rangle+
 \langle\hat{J}_{2i}^a\rangle+ \langle\hat{J}_{2i+1}^a\rangle=0
\end{equation}
as well as 
\begin{equation}
 C_{i,J_j^b}^a :=
 \langle(\hat{J}_j^b-\langle\hat{J}_j^b\rangle)\hat{C}_i^a\rangle=  
-\Delta(J_j^bJ_{2i-1}^a)+ \Delta(J_j^bJ_{2i}^a)+ \Delta(J_j^bJ_{2i+1}^a)=0
\end{equation}
and higher-order constraints.

For the boundary expectation values, the equations of motion are
\begin{equation}
\label{expect1X}
\frac{{\rm d}\langle\hat{J}_{1}^a\rangle}{{\rm d} t} = \alpha \,
\epsilon^{abc} \, 
\big(\langle\hat{J}_{3}^b\rangle \langle\hat{J}_{1}^c\rangle + \Delta(J_{3}^b
J_{1}^c) - \langle\hat{J}_{2}^b\rangle \langle\hat{J}_{1}^c\rangle -
\Delta(J_{2}^b  J_{1}^c) \big) 
\end{equation}
and
\begin{equation}
\frac{{\rm d}\langle\hat{J}_{2N+1}^a\rangle}{{\rm d} t} = \alpha \,
\epsilon^{abc} \, 
\big(\langle\hat{J}_{2N}^b\rangle \langle\hat{J}_{2N+1}^c\rangle +
\Delta(J_{2N}^b J_{2N+1}^c) - \langle\hat{J}_{2N-1}^b\rangle 
\langle\hat{J}_{2N+1}^c\rangle - \Delta(J_{2N-1}^b J_{2N+1}^c) \big)\,.
\end{equation}
Implementing the constraint, we obtain
\begin{equation}
\label{expect2X}
\frac{{\rm d}\langle\hat{J}_{1}^a\rangle}{{\rm d} t} = 2 \, \alpha \, \epsilon^{abc} \,
\big( \langle\hat{J}_{3}^b\rangle \langle\hat{J}_{1}^c\rangle + \Delta(J_{3}^b J_{1}^c) \big) 
\end{equation}
and
\begin{equation}
\frac{{\rm d}\langle\hat{J}_{2N+1}^a\rangle}{{\rm d} t} = 2 \, \alpha \,
\epsilon^{abc} \, 
\big( \langle\hat{J}_{2N-1}^b\rangle \langle\hat{J}_{2N+1}^c\rangle  +
\Delta(J_{2N-1}^b J_{2N+1}^c) \big) \,. 
\end{equation}
The reduced equations for internal spins are
\begin{equation} \label{Jt}
\frac{{\rm d}\langle\hat{J}_{2i+1}^a\rangle}{{\rm d} t} =  2 \, \alpha
\, \epsilon^{abc} 
\, \big(\langle\hat{J}_{2i-1}^b\rangle \langle\hat{J}_{2i+1}^c\rangle +
\Delta(J_{2i-1}^b J_{2i+1}^c) + 
 \langle\hat{J}_{2i+3}^b\rangle \langle\hat{J}_{2i+1}^c\rangle +
 \Delta(J_{2i+3}^b J_{2i+1}^c) \big)  
\end{equation}
where $i=\{2, \cdots, N-1\}$.  The equation of motion generated by the
effective version of the reduced quantum Hamiltonian (\ref{H_Qred}) are
\begin{equation}\label{Jt2}
\frac{{\rm d}\langle\hat{J}_{2i+1}^{\mathrm{red},a}\rangle}{{\rm d} t} =
3 \, \alpha \, 
\epsilon^{abc} \, \big(\langle\hat{J}_{2i-1}^b\rangle
\langle\hat{J}_{2i+1}^c\rangle + \Delta(J_{2i-1}^b 
 J_{2i+1}^c) + \langle\hat{J}_{2i+3}^b\rangle \langle\hat{J}_{2i+1}^c\rangle +
 \Delta(J_{2i+3}^b J_{2i+1}^c) \big) \,. 
\end{equation}
As in the case of operator equations, the equations can be mapped into each
other by a constant rescaling of the time coordinate. We note that these
equations are exact as no truncation by moments has been necessary. However,
equations of motion for second-order moments depend on third-order moments and
have to be truncated for a self-contained semiclassical approximation to
first order in $\hbar$.

Analytical solutions of these non-linear equations for large $N$ are difficult
to find. However, it is possible to analyze some general properties of
interest in the context of long-range correlations.  In particular, we are
interested in correlations between spins at the two boundaries of the graph,
which are certainly long-range for large $N$. We will assume that initially
there are no correlations between different spins, but they will build up over
time as the system evolves.

The reduced equation of motion for the correlation $\Delta(J_{1}^a
J_{2i+1}^b)$ is
\begin{eqnarray}
\label{coupled}
\frac{{\rm d} \Delta(J_1^a J_{2i+1}^b)}{{\rm d} t} &=& 2 \, \alpha \,
\bigg\{ \epsilon^{acd} \, \big(\langle\hat{J}_1^d\rangle \Delta(J_3^c
J_{2i+1}^b) + \langle\hat{J}_3^c\rangle 
\Delta(J_1^d J_{2i+1}^b \big) \nn \\ 
&&+ \epsilon^{bef} \big(\langle\hat{J}_{2i-1}^e\rangle \Delta(J_1^a
J_{2i+1}^f) + \langle\hat{J}_{2i+1}^f\rangle 
\Delta(J_1^a J_{2i-1}^e)\big) \bigg\} \,.
\end{eqnarray}
If we start with an uncorrelated state, the initial conditions are such that
the only non-zero initial values are the expectation value of $\hat{J}_1$ and
$\hat{J}_{2N+1}$ and their variances chosen such that uncertainty relations
are respected. For interior spins, only the fluctuations are assumed non-zero
(although they would be allowed to be zero for a general spin system).

\begin{figure}[htbp]
\begin{center}
 \includegraphics[scale=1]{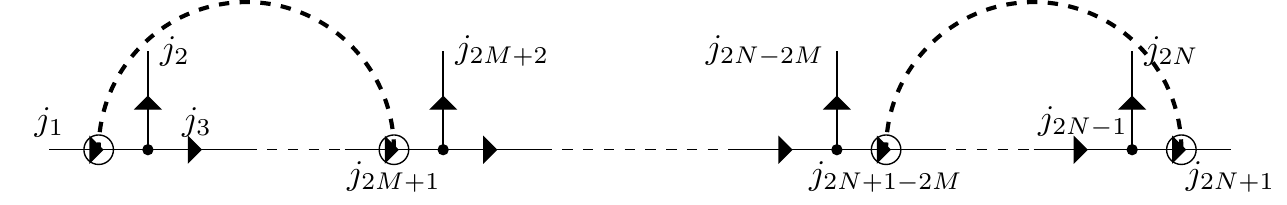}
\end{center}
\caption{Correlations between the farthest spin sets for $\hat{\vv J}_1$ and
  $\hat{\vv J}_{2N+1}$ after solving the equations $M$ times iteratively for a
  graph with $N$ vertices for $2M \leq N +1$. The semi-circles represent
  non-zero covariances $\Delta(J_{1}^a J_{2M+1}^b)$ and $\Delta(J_{2N+1}^a
  J_{2N+1-2M}^b)$. \label{Fig:CorrM}}
\end{figure}

\begin{figure}[htbp]
\begin{center}
 \includegraphics[scale=1]{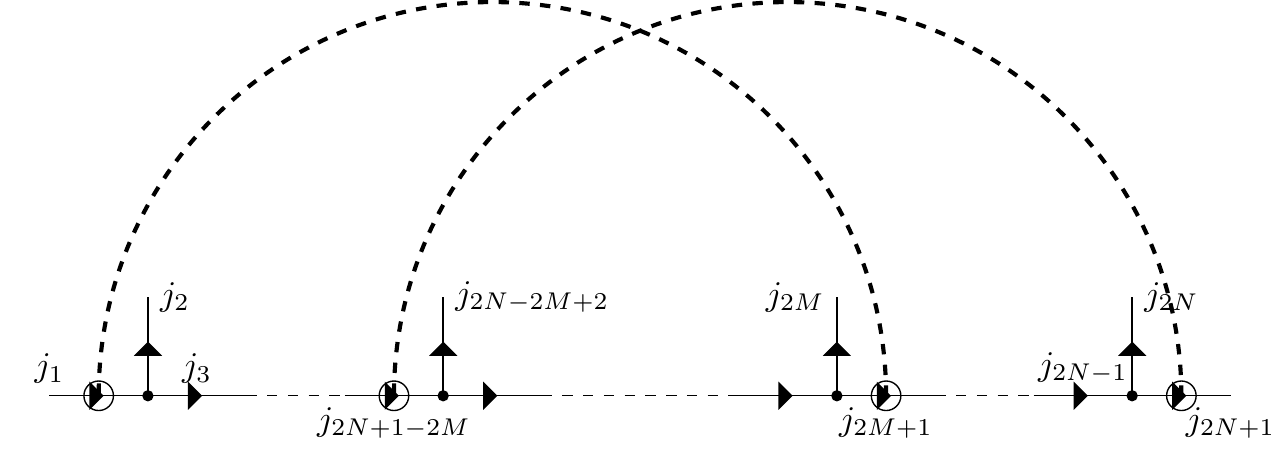}
\end{center}
\caption{Correlations between the farthest spin sets for $\hat{\vv J}_1$ and
  $\hat{\vv J}_{2N+1}$ after solving the equations $M$ times iteratively for a
  graph with $N$ vertices for $N < 2M$. The semi-circles represent non-zero
  covariances $\Delta(J_{1}^a J_{2M+1}^b)$ and $\Delta(J_{2N+1}^a
  J_{2N+1-2M}^b)$. \label{Fig:Corr2M}}
\label{fig:3}
\end{figure}

The expectation values and second-order moments of the state for all spins are
coupled to one another through the evolution equations.  One may solve them
perturbatively by orders of $\alpha$. To zeroth order, all expectation values
and moments are constant and no correlations build up.  To first order, using
(\ref{coupled}), ${\rm d} \Delta(J_1J_3)/{\rm d}t$ has a non-zero contribution
of the form $\alpha \langle\hat{J}_1\rangle \Delta(J_3J_3)$ and after some
time $\Delta(J_1J_3)$ is non-zero to first order in $\alpha$. The expectation
value $\langle\hat{J}_3\rangle$ has a time derivative of the form (\ref{Jt})
or (\ref{Jt2}) with one non-zero term of the form $\alpha \Delta(J_1J_3)$ and
after some time is non-zero to second order in $\alpha$. We can iterate this
procedure and generate non-zero expectation values on all links, as well as
non-zero correlations between the links. The first step has generated a
covariance $\Delta(J_1J_3)=\mathcal{O}(\alpha\hbar)$ because we used one solution
perturbative in $\alpha$ and one fluctuation. For the next step, moving up to
$\Delta(J_1J_5)$, we first need to generate a non-zero
$\langle\hat{J}_3\rangle$ and $\Delta(J_3J_5)$, using one additional
fluctuation and repeated perturbative solutions. We therefore obtain a
non-zero $\Delta(J_1J_5)=\mathcal{O}(\alpha^4\hbar^2)$. By iteration, we obtain a
non-zero $\Delta(J_1J_{2i+1})=\mathcal{O}(\alpha^{3i-2}\hbar^i)$. For long-range
correlations between the boundary spins, we need to apply the procedure up to
$i=N/2$, after which all spins are correlated from non-zero seed expectation
values on both boundaries; see Figs.~\ref{Fig:CorrM} and \ref{Fig:Corr2M}. A
small number of iterations, $M \ll N$, does not lead to strong long-range
correlations. The building-up of long-range correlations from local dynamics
can therefore be seen only to high perturbative orders, or non-perturbatively,
and it requires high orders in an $\hbar$-expansion.

\subsection{Dynamical stability for small number of vertices}
\label{s:Small}

We now return to the question of dynamical stability of states of the spin
chain as an isolated system. We will consider discrete theories with small
number of vertices and choose initial values corresponding to the different
ground states found in our minisuperspace models. The effective evolution
equations can be solved perturbatively and indicate dynamical stability in
some cases. (We will not analyze stability under perturbations of initial
configurations.) Although we present explicit solutions only for small numbers
of vertices, the relevant features are generic and can be seen also for large
numbers of vertices, but it is more cumbersome to produce explicit
expressions.

\subsubsection{Single-vertex graph}

For $N=1$, we have two horizontal spins which we can identify with the spins
in our two interacting minisuperspace models, or view as a theory beyond the
minimal minisuperspace model.  In this case, correlations between the boundary
spins are next-neighbour correlations. The reduced equations of motion are
\begin{equation}
\label{expect1c}
\frac{{\rm d}\langle\hat{J}_{1}^x\rangle}{{\rm d} t} = 2 \, \alpha \,
\big(\langle\hat{J}_{1}^z\rangle \langle\hat{J}_{3}^y\rangle - 
\langle\hat{J}_{1}^y\rangle \langle\hat{J}_{3}^z\rangle + \Delta(J_{1}^z
J_{3}^y) - \Delta(J_{1}^y J_{3}^z) \big)  
\end{equation}
and cyclic for the $y$ and $z$ components. For the second
horizontal spin, $\vv J_3$, we have
\begin{equation}
\label{sol12}
\frac{{\rm d}\langle\hat{J}_{3}^x\rangle}{{\rm d} t} = 2 \, \alpha \,
\big(\langle\hat{J}_{1}^y\rangle \langle\hat{J}_{3}^z\rangle - 
\langle\hat{J}_{1}^z\rangle \langle\hat{J}_{3}^y\rangle + \Delta(J_{1}^y
J_{3}^z) - \Delta(J_{1}^z J_{3}^y)\big) = - 
\frac{{\rm d}\langle\hat{J}_{1}^x\rangle}{{\rm d} t} 
\end{equation}
and again cyclic for the $y$ and $z$ components. The equality ${\rm
  d}\langle\hat{\vv J}_1\rangle/{\rm d}t = -{\rm d}\langle\hat{\vv
  J}_3\rangle/{\rm d}t$ can be read off from the equations of motion, but it
also follows directly from the conservation of (\ref{G}). The vertical spin
can be obtained from the horizontal ones by the simple constraint
$\langle\hat{\vv J}_2\rangle =\langle\hat{\vv J}_1\rangle -\langle\hat{\vv
  J}_3\rangle$.

We will first solve the corresponding classical equations, which are as above
but with zero covariances. We proceed by perturbation theory with respect to
$\alpha$, so that we have constant spins to zeroth order. Upon repeatedly
inserting lower-order solutions in the equations of motion, we proceed up to
second order and obtain
\begin{eqnarray} \label{Jclass}
\vv J_1^{(2)}(t) = \vv A_{1}  +  2 \, \alpha \, \vv B_1 \,
t + 4 \, \alpha^2 \, \vv C_1 t^2 +\cdots \\ 
\vv J_3^{(2)}(t) = \vv A_{3}  + 2 \, \alpha \, \vv B_3 \,
t + 4 \,  \alpha^2 \, \vv C_3 t^2 +\cdots \, ,
\end{eqnarray}
where $\vv A_{i}$ are free constant vectors.  These approximate solutions are
valid up to $t \sim 1/\alpha $, respecting the perturbative regime.

The remaining coefficients
\begin{equation} \label{B}
 B_1^x = A_1^zA_3^y-A_1^yA_3^z
\end{equation}
and 
\begin{equation} \label{C}
 C_1^x= -A_1^yB_3^z-B_1^yA_3^z+A_1^zB_3^y+B_1^zA_3^y\,,
\end{equation}
and cyclic for the $y$ and $z$ components, are also constant but strictly
related to $\vv A_i$. From (\ref{sol12}), we have that $\vv B_1=-\vv B_3$ and
$\vv C_1=-\vv C_3$. Therefore,
\begin{eqnarray}
 C_1^x&=& (A_1^zA_3^x-A_1^xA_3^z) A_3^z-A_1^y (A_1^xA_3^y-A_1^yA_3^x) -
 (A_1^xA_3^y-A_1^yA_3^x) A_3^y+A_1^z (A_1^zA_3^x-A_1^xA_3^z)\nonumber\\
 &=& -A_1^x \left((A_3^y)^2+ (A_3^z)^2\right) + A_3^x \left((A_1^y)^2+
   (A_1^z)^2\right) + A_1^zA_3^z \left(A_3^x-A_1^x\right) 
+ A_1^yA_3^y \left(A_3^x-A_1^x\right) \nonumber \\
&=& A_3^x \, |\vv A_1|^2 - A_1^x \, |\vv A_3|^2 + (\vv A_1 \cdot \vv A_3) 
\, (A_3^x - A_1^x) \,, \label{C2}
\end{eqnarray}
which enables the vector coefficient $\vv C_3$ expressing purely in
terms of constant vectors $\vv A_i = A_i \hat{A}_i$:
\begin{eqnarray} \label{C_1}
\vv C_3 = -\vv C_1 &=& A_3^2 \vv A_1 - A_1^2 \vv A_3 + (\vv A_1 \cdot 
\vv A_3) \, (\vv A_1 - \vv A_3) \nn \\
&=& A_1 A_3\big[(A_3 + (\hat{A}_1 \cdot \hat{A}_3) A_1) \hat{A}_1 - 
(A_1 + (\hat{A}_1 \cdot \hat{A}_3) {A}_3) \hat{A}_3\big] \, .
\end{eqnarray} 
As one possible choice of initial conditions, we could impose that all spins
other than the boundary ones (that is, only the vertical spin in the present
model) are zero. Therefore, for a single vertex, $\vv J_2(0)=0$ and $\vv
A_1 = \vv A_3$. Equations~(\ref{B}) and (\ref{C}) then imply that
$\vv B_i=0=\vv C_i$, and all spins remain constant in time. This result,
although it is classical, agrees with the trivial dynamics in our minimal
minisuperspace model.

We now include moment terms and find solutions of the quantum theory, again
perturbative in $\alpha$.  In order to obtain information about the boundary
correlations, we should compute quantities such as $\Delta(J_{1}^x J_{3}^y)$
from
\begin{eqnarray} \label{dD1}
\frac{{\rm d} \Delta(J_{1}^x J_{3}^y)}{{\rm d} t} = 2 \, \alpha \,
\big(-\langle\hat{J}_{1}^y\rangle \Delta(J_{3}^y J_{3}^z) +
\langle\hat{J}_{3}^y\rangle \Delta(J_{1}^z J_{3}^y) +\langle\hat{J}_{1}^z\rangle 
\Delta((J_{3}^y)^2) - \langle\hat{J}_{3}^z\rangle \Delta(J_{1}^y J_{3}^y) \nn \\ 
+ \langle\hat{J}_{1}^z\rangle \Delta(J_{1}^x J_{3}^x) -
\langle\hat{J}_{1}^x\rangle  \Delta(J_{1}^x J_{3}^z) - \langle\hat{J}_{3}^z\rangle
\Delta((J_{1}^x)^2) + \langle\hat{J}_{3}^x\rangle \Delta(J_{1}^x J_{1}^z)\big) \,.
\end{eqnarray}
We will also need to consider
 \begin{eqnarray} \label{dD2}
\frac{{\rm d} \Delta(J_{1}^y J_{3}^z)}{{\rm d} t} = 2 \, \alpha \,
\big(-\langle\hat{J}_{1}^z\rangle \Delta(J_{3}^z J_{3}^x) +
\langle\hat{J}_{3}^z\rangle \Delta(J_{1}^x J_{3}^z) +\langle\hat{J}_{1}^x\rangle 
\Delta((J_{3}^z)^2) - \langle\hat{J}_{3}^x\rangle \Delta(J_{1}^z J_{3}^z) \nn \\ 
+ \langle\hat{J}_{1}^x\rangle \Delta(J_{1}^y J_{3}^y) -
\langle\hat{J}_{1}^y\rangle  \Delta(J_{1}^y J_{3}^x) - \langle\hat{J}_{3}^x\rangle
\Delta((J_{1}^y)^2) + \langle\hat{J}_{3}^y\rangle \Delta(J_{1}^y J_{1}^x)\big) \,.
\end{eqnarray}
and
 \begin{eqnarray} \label{dD3}
\frac{{\rm d} \Delta(J_{1}^x J_{3}^z)}{{\rm d} t} = 2 \, \alpha \,
\big(-\langle\hat{J}_{1}^y\rangle \Delta((J_{3}^z)^2) +
\langle\hat{J}_{3}^y\rangle \Delta(J_{1}^z J_{3}^z) +\langle\hat{J}_{1}^z\rangle 
\Delta(J_{3}^yJ_3^z) - \langle\hat{J}_{3}^z\rangle \Delta(J_{1}^y J_{3}^z) \nn \\ 
+ \langle\hat{J}_{1}^x\rangle \Delta(J_{1}^x J_{3}^y) -
\langle\hat{J}_{1}^y\rangle  \Delta(J_{1}^x J_{3}^x) - \langle\hat{J}_{3}^x\rangle
\Delta(J_1^xJ_{1}^y) + \langle\hat{J}_{3}^y\rangle \Delta((J_{1}^x)^2)\big) \,.
\end{eqnarray}

For generic initial conditions, these covariances will have the same quadratic
form to second order in $\alpha$, using non-zero initial
fluctuations. However, in some specific cases the covariances remain constant,
corresponding to stable initial configurations. In particular, we are
interested in whether our classical solutions (\ref{Jclass}) are
perturbatively stable within a semiclassical treatment of the quantum
dynamics. We must then test whether the covariance terms in (\ref{expect1c})
change the behavior.

As before, we first assume that the initial expectation values are such that
$\vv J_2(0)=0$, or $\vv A_1=\vv A_3$. Moreover, we assume fluctuations and
covariances as we found them for the corresponding ground state in the minimal
minisuperspace model, given by (\ref{Deltaxx}):
$\Delta((J^x)^2)=\Delta((J^y)^2)=\frac{1}{2}\hbar
|\langle\hat{J}^z\rangle|$, now for both horizontal spins in the
single-vertex model. This result had been derived by assuming the spin
expectation values to point in the $z$-direction, which we will also do
now. Moreover, we have initially zero covariances between components of the
two spins.

Assuming the spin expectation values to point in the $z$-direction leaves only
three non-zero terms in (\ref{dD1}), two of which vanish for zero initial
covariances. We are left with $\langle\hat{J}_1^z\rangle \Delta((J_3^y)^2)-
\langle\hat{J}_3^z\rangle \Delta((J_1^y)^2)$. This difference is zero
initially because the fluctuations and expectation values on the two
horizontal links are the same. Therefore, ${\rm d}\Delta(J_1^xJ_3^y)/{\rm
  d}t=0$ and this covariance remains zero to the orders considered
here. Similarly, (\ref{dD2}) and (\ref{dD3}) remain zero, and the covariance
terms in (\ref{expect1c}) do not contribute for this choice of initial
values. The configuration corresponding to the minimal minisuperspace model is
therefore dynamically stable within the single-vertex model.

For the ground-state configurations of the interacting minisuperspace models
we also obtain perturbative stability, but the arguments are slightly
different in the case of antiparallel alignment. In (\ref{dD1}), the
fluctuation terms no longer cancel out because
$\langle\hat{J}_3^z\rangle=-\langle\hat{J}_1^z\rangle$. However, there are now
two non-zero covariance terms in (\ref{dD1}) because
$\Delta(J_1^xJ_3^x)=\Delta(J_1^yJ_3^y)= -\frac{1}{2}
\hbar|\langle\hat{J}^z\rangle|$ from (\ref{Delta12}), where
$\langle\hat{J}^z\rangle$ on the right could now refer to either $\hat{J}_1^z$
or $\hat{J}_3^z$ because their absolute values are equal. We now have four
non-zero individual terms in (\ref{dD1}), but they all cancel out for the
given initial values. Again, we have dynamical stability of the ground
state.

Before we move on to two vertices, we confirm the ground-state covariances for
a spin-$1/2$ system. For $\gamma<0$ and spins pointing in the $z$-direction,
the ground state $|J_1^z,J_3^z\rangle=|1/2,1/2\rangle$ is uncorrelated and has
zero spin-spin covariance $\Delta(J_1^xJ_3^x)$ as used. For $\gamma<0$, the
ground state is the singlet $2^{-1/2} (|1/2,-1/2\rangle-|-1/2,1/2\rangle$,
which is correlated and leads to
$\langle\hat{J}_1^x\hat{J}_3^x\rangle=-\frac{1}{4}\hbar^2$ by standard
calculations. Since $\langle\hat{J}_1^x\rangle=0=\langle\hat{J}_3^x\rangle$ in
this state, we have $\Delta(J_1^xJ_3^x)=-\frac{1}{4}\hbar^2= -\frac{1}{2}\hbar
|\langle\hat{J}_3^z\rangle|$ as derived in (\ref{Delta12}).

\subsubsection{Two-vertex graph and beyond}

\begin{center}
\begin{figure}[htbp]
\includegraphics[scale=1]{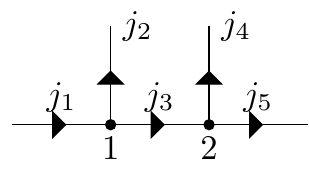}
\caption{A one-dimensional two-vertex graph.}
\end{figure}
\end{center}

For the two-vertex graph, the evolution equations of the boundary spins at
the edges are of a similar form as in the case of a single-vertex graph:
\begin{eqnarray} 
\frac{{\rm d}\langle\hat{J}_{1}^x\rangle}{{\rm d} t} = 2 \, \alpha \,
\big(\langle\hat{J}_{1}^z\rangle \langle\hat{J}_{3}^y\rangle  -
\langle\hat{J}_{1}^y\rangle \langle\hat{J}_{3}^z\rangle + \Delta(J_{1}^z
J_{3}^y) - \Delta(J_{1}^y J_{3}^z)\big) \\  
\frac{{\rm d}\langle\hat{J}_{5}^x\rangle}{{\rm d} t} = 2 \, \alpha \,
\big(\langle\hat{J}_{3}^y\rangle \langle\hat{J}_{5}^z\rangle  -
\langle\hat{J}_{3}^z\rangle \langle\hat{J}_{5}^y\rangle + \Delta(J_{3}^y
J_{5}^z) - \Delta(J_{3}^z J_{5}^y)\big)  
\end{eqnarray}
but now they couple to internal spins.
For the internal horizontal spin, $\hat{\vv J}_3$, we have 
\begin{eqnarray}
\frac{{\rm d}\langle\hat{J}_{3}^x\rangle}{{\rm d} t} &=& 2 \, \alpha \,
\big((\langle\hat{J}_{1}^y\rangle + \langle\hat{J}_{5}^y\rangle) 
\langle\hat{J}_{3}^z\rangle - (\langle\hat{J}_{1}^z\rangle +
\langle\hat{J}_{5}^z\rangle) \langle\hat{J}_{3}^y\rangle  + \Delta(J_{1}^y
J_{3}^z) - 
\Delta(J_{1}^z J_{3}^y) +  \Delta(J_{3}^z J_{5}^y) - \Delta(J_{3}^y
J_{5}^z)\big)\nonumber\\
&=& - \frac{{\rm d}\langle\hat{J}_{1}^x\rangle}{{\rm d} t}-\frac{{\rm
    d}\langle\hat{J}_{5}^x\rangle}{{\rm     d} t}  \,. \label{J315}
\end{eqnarray}
Classical solutions up to second order in $\alpha$ are
\begin{eqnarray}
\vv J_1^{(2)}(t) &=& \vv A_1 + 2 \, \alpha \, \vv B_1 \, t + 4 \,
\alpha^2 \, \vv C_1 t^2 +\cdots \\ 
\vv J_5^{(2)}(t) &=&  \vv A_5+ 2 \, \alpha \, \vv B_5 \, t + 4 \,
\alpha^2 \, \vv C_5  t^2 +\cdots\,.
\end{eqnarray}
With (\ref{J315}), the solution for the internal spin is
\begin{equation}
\vv J_3^{(2)}(t) = \vv A_3 + 2 \, \alpha \, \vv B_3 \,
t + 4 \, \alpha^2 \, \vv C_3 t^2+\cdots \,,
\end{equation}
where $\vv B_3 = -(\vv B_1 + \vv B_5)$ and $\vv C_3 = -(\vv C_1 + \vv C_5)$ follow from the conversation of the total spin. The coefficients
$\vv C_3$ can be obtained by replacing $\vv A_1 \rightarrow \vv A_1 + \vv A_5$
in (\ref{C_1}):
\begin{equation} \label{C_3} 
\vv C_3 =
A_3^2 \, (\vv A_1 + \vv A_5) - | \vv A_1 + \vv A_5 |^2 \vv A_3 + \big[(\vv A_1
+ \vv A_5) \cdot \vv A_3\big] \, (\vv A_1 + \vv A_5 - \vv A_3) \, .  
\end{equation}
For the vertical spins $\vv J_2$ and $\vv J_4$ we then have
\begin{eqnarray}
\vv J_2^{(2)}(t) &=& \vv J_1^{(2)}(t)-\vv J_3^{(2)}(t) = \vv A_1-\vv A_3+ 2 \,
\alpha \, (2\vv B_1 +   \, \vv B_5) 
\, t + 4 \, \alpha^2 \, (2\vv C_1 + \vv C_5) t^2+\cdots \\ 
\vv J_4^{(2)}(t) &=& \vv J_3^{(2)}(t)-\vv J_5^{(2)}(t) = \vv A_3-\vv A_5- 2 \,
\alpha \, (\vv B_1 +  2 \, \vv B_5) 
\, t -  4 \, \alpha^2 \, (\vv C_1 + 2\vv C_5) t^2+\cdots \,.
\end{eqnarray}

We need equations of motion for covariances in order to extend the classical
solutions to the semiclassical regime. These equations for correlations of
neighboring spins are very similar to the equations of the single-vertex case,
but we can now also have changing values of more distant spins, such as
\begin{eqnarray} \label{dJ15dt}
  \frac{{\rm d} \Delta(J_{1}^x J_{5}^y)}{{\rm d} t} &=& 2 \, \alpha \,
  \big(-\langle\hat{J}_{1}^y\rangle 
  \Delta(J_{3}^z J_{5}^y) + \langle\hat{J}_{3}^y\rangle \Delta(J_{1}^z J_{5}^y)
  + \langle\hat{J}_{1}^z\rangle  \Delta(J_{3}^y
  J_{5}^y) - \langle\hat{J}_{3}^z\rangle \Delta(J_{1}^y J_{5}^y) \nn \\ 
  &&- \langle\hat{J}_{3}^x\rangle \Delta(J_{1}^x J_{5}^z) -
  \langle\hat{J}_{5}^z\rangle  \Delta(J_{1}^x J_{3}^x) +
  \langle\hat{J}_{3}^z\rangle 
  \Delta(J_{1}^x J_{5}^x) + \langle\hat{J}_{5}^x\rangle \Delta(J_{1}^x
  J_{3}^z)\big)  
\end{eqnarray}
for the boundary spins of the two-vertex graph. For spin expectation values
pointing in the $z$-direction, there are four potentially non-zero terms,
$\langle\hat{J}_{1}^z\rangle \Delta(J_{3}^y J_{5}^y)-
\langle\hat{J}_{5}^z\rangle \Delta(J_{1}^x J_{3}^x)+
\langle\hat{J}_{3}^z\rangle (\Delta(J_{1}^x J_{5}^x)-\Delta(J_{1}^y
J_{5}^y))$. The covariances are zero unless we have a state with antiparallel
orientation of neighboring spins. The four remaining terms then cancel out
because $\langle\hat{J}_{1}^z\rangle= -\langle\hat{J}_{3}^z\rangle=
\langle\hat{J}_{5}^z\rangle$ and $\Delta(J_{1}^x J_{5}^x)=\Delta(J_{1}^y
J_{5}^y)$. 

At this point we have to be careful when we compare minisuperspace
configurations with the exact ground state(s) of the two-vertex model, which
is an odd-number spin chain and has strong finite-size effects. Diagonalizing
the Hamiltonian $\alpha(\hat{\vv J}_1\cdot\hat{\vv J}_3+ \hat{\vv J}_3\cdot
\hat{\vv J}_5)$ in the spin-$1/2$ case leads to the degenerate ground states
$\psi_1= 6^{-1/2} (|-1/2,1/2,1/2\rangle- 2|1/2,-1/2,1/2\rangle+
|1/2,1/2,-1/2\rangle)$ and $\psi_2= 6^{-1/2} (|-1/2,-1/2,1/2\rangle-
2|-1/2,1/2,-1/2\rangle+|1/2,-1/2,-1/2\rangle)$. Choosing the first state to be
specific, one can then compute the expectation values
$\langle\hat{J}_1^z\rangle= \frac{1}{3}\hbar= \langle\hat{J}_5^z\rangle$ and
$\langle\hat{J}_3^z\rangle= -\frac{1}{6}\hbar$. These values are next-neighbor
antiparallel, but do not obey $\langle\hat{J}_{1}^z\rangle=
-\langle\hat{J}_{3}^z\rangle= \langle\hat{J}_{5}^z\rangle$. Moreover, we have
the covariances $\Delta(J_1^xJ_3^x)=-\frac{1}{6}\hbar^2=\Delta(J_1^yJ_3^y)$
which do not obey (\ref{Delta12}), and we have
$\Delta(J_1^zJ_3^z)=-\frac{1}{9}\hbar^2$. There are also distant covariances
such as $\Delta(J_1^xJ_5^x)=\frac{1}{12}\hbar^2=\Delta(J_1^yJ_5^y)$ and
$\Delta(J_1^zJ_5^z)=-\frac{1}{36}\hbar^2$. Although these values do not show
the generic antiparallel behavior, one can still see that all terms in
(\ref{dJ15dt}) cancel out.

The expressions for $\vv B_i$ and $\vv C_i$ in terms of $\vv A_i$ are very
similar to those in (\ref{B}) and (\ref{C}), just with different labels. (The
relation (\ref{C2}) for $\vv C_i$ in terms of $\vv A_i$, however, has a
different form for multiple vertices because it has been derived for the
single-vertex graph using $\vv B_3=-\vv B_1$.)  For ground-state
configurations of minisuperspace models we therefore obtain the same
cancelations as in the single-vertex model because these considerations depend
only on the expectation values and moments of horizontal spins on neighboring
links. In the two-vertex model, one can choose initial conditions such that
all vertical spins have zero expectation values and covariances with any other
spins. Therefore, $\vv J_4^{(2)}(0) = \vv J_2^{(2)}(0)=0$ and $\vv A_1^{(0)} =
\vv A_3^{(0)} = \vv A_5^{(0)}$. This implies that $\vv B_i=0=\vv C_i$, and all
spins remain constant in time as in the single-vertex.

The same pattern is then realized also for more vertices, and we conclude that
the various ground states are dynamically stable. The minisuperspace models
with Hamiltonians $\hat{H}_{\rm mini}^{(1)}$ and $\hat{H}_{\rm mini}^{(2)}$
predict the same ground-state configurations as the discrete theory, and these
minisuperspace states are therefore stable within the discrete
model. The minisuperspace model with Hamiltonian $\hat{H}_{\rm mini}^{(3)}$,
however, is unstable. It not only predicts a ground-state configuration that
does not agree with any fundamental ground state, it also has unstable
dynamics when its ground-state configuration is embedded in a fundamental spin
chain. Unlike in the other two models, there are then three neighboring spins
$\vv J_{2i_{\rm c}-3}$, $\vv J_{2i_{\rm c}-1}$ and $\vv J_{2i_{\rm c}+1}$ around the
central vertex at $i_{\rm c}=(N+1)/2$, such that $\langle\hat{\vv J}_{2i_{\rm
    c}-3}\rangle= \langle\hat{\vv J}_{2i_{\rm c}-1}\rangle= -\langle\hat{\vv
  J}_{2i_{\rm c}+1}\rangle$, and no cancellations happen in (\ref{dJ15dt}).

\section{Possible implications for quantum cosmology}
\label{s:QC}

We have analyzed four different minisuperspace models of a spin system
related to the Heisenberg spin chain. The first model resembles the
traditional construction of quantum-cosmology models in which only
homogeneous degrees od freedom without spatial variation are
considered. In the present context, such a model does not capture the
dynamical nature of coupled spins. The second model is a more recent
proposal to apply mathematical constructions of condensate states to
quantum cosmology. The dynamics is then quite different from a
traditional minisuperspace model. In particular, a non-trivial
dynamics is now realized even though the degrees of freedom included
in the model are the same. The remaining models incorporate additional
degrees of freedom in two different ways. They both lead to
non-trivial dynamics and in this sense improve the reduction. However,
details of ground states and stability are very different in the two
models, indicating that good knowledge of the fundamental dynamics is
important for a successful construction of minisuperspace models. This
conclusion is our first result in an application to quantum cosmology:
Traditional minisuperspace models start with a reduction of the
classical theory, and then quantize by using some ideas related to
candidates for full quantum gravity. But if they do not directly
address the full dynamics, they may be in danger of missing crucial
information, just as our model Hamiltonian $\hat{H}_{\rm mini}^{(4)}$
does compared with $\hat{H}_{\rm mini}^{(3)}$.

Our successful Hamiltonian $\hat{H}_{\rm mini}^{(3)}$ is similar to the
sublattice Hamiltonians introduced in \cite{FiniteHeisenberg} for Heisenberg
spin chains. The detailed analysis of this paper showed that such an
approximation, for given chain length $N$, is better for larger spin lengths
$s$ on the chain. If a similar statement is true for the dynamics of quantum
cosmology, it would indicate that a coarse-graining procedure applied before
symmetry reduction could improve models of quantum cosmology, for such a
procedure would combine the small fundamental spins of $s=1/2$ to systems of
larger spin lengths. Coarse graining in gravitational, and in particular
background-independent theories, is not well-understood, but proposals have
been made for instance in \cite{Nets,CoarseGrainingFlow}. It is also
encouraging that sublattice structures can be found in spin systems with more
than one dimensions \cite{Sublattice,Sublattice2}.

Our minisuperspace Hamiltonian $\hat{H}_{\rm mini}^{(4)}$ is based on a
construction similar to the separate-universe approximation of classical
cosmology. Its failure to model properties of ground states and stability
indicates that it is not a good quantum approximation for all kinds of
fundamental dynamics. Its lack of stability is of particular concern,
resulting from the fact that in this model the minisuperspace ground state
does not correspond to the ground state of the discrete theory. We therefore
have provided an explicit example of important fundamental properties not
captured by a minisuperspace model. Such models are unstable if energy can be
exchanged with an environment, and one might conclude that they are
unreliable.

However, not just ground states but also excited states may be stable
in an isolated system if no energy can be exchanged with an
environment. This is the situation usually realized in models of
quantum cosmology, where the state represents the whole universe with
nothing outside. The stability of excited states then results in a
large variety of candidates for homogeneous
configurations. Nevertheless, some caution toward such minisuperspace
states used in quantum cosmology is still required: Our spin system
can be taken as a model for quantum space, in which near homogeneity
should be possible and stable under evolution. However, there should
also be matter, with additional degrees of freedom that could be
placed on the same graph used for our spin chain but representing a
different system of degrees of freedom. There could then be energy
exchange between the spin system analyzed here and the new matter
system. We would be back at the question whether a homogeneous spin
configuration, representing quantum space, can be stable within the
coupled system if it does not capture the correct ground state. The
question of how matter is coupled to quantum space therefore seems
important in the context of the emergence and stability of correlated
quantum-cosmology states.

\section{Discussion}

We have analyzed a discrete spin model with different methods 
used in recent years in 
canonical quantum gravity. Our aim is to test the latter, rather than
revealing new properties of spin systems in general. We have found new results
in three different classes: minisuperspace truncations, effective theories,
and dynamical long-range correlations.

We have derived different minisuperspace models directly from the discrete
theory, which is a new procedure compared with the usual construction of
minisuperspace models by quantizing homogeneous configurations of a continuum
theory. Several novel features could be seen, for instance the existence of
different minisuperspace models of the same discrete theory, paralleling the
existence of different continuum limits of one discrete theory. In our
specific constructions, starting with the discrete theory has the advantage
that no problems related to infrared scales of traditional minisuperspace
models occur. We have seen that it can be of advantage to keep more degrees of
freedom in a minisuperspace model than simple homogeneous configurations would
suggest, in particular when non-trivial dynamical properties should be
obtained. Good knowledge of the fundamental theory is required in order to
select a reliable minisuperspace model. Alternatively, non-trivial dynamics
can be obtained by using condensate states, as employed also in cosmological
models of group-field theories
\cite{GFTCosmo,GFTCosmo2,GFTCosmo3,GFTLattice,GFTPerturb,GFTLQC,GFTCyclic,GFTImpact}.

In several examples of our minisuperspace models, we have computed canonical
effective potentials and equations and found good agreement with known
ground-state energies and configurations. Our results provide further support
for the canonical effective methods proposed for quantum gravity in
\cite{EffAc,Karpacz}, with an extension to the computation of effective
potentials in \cite{CW}.

We have also analyzed the discrete spin system directly, with an emphasis on
properties that should be important for the dynamical building-up of
long-range correlations as they are likely to be relevant for the dynamical
emergence of states that may be described by minisuperspace models, a
question related to the continuum limit of discrete quantum gravity. Our
analysis, based on rather general properties of the underlying equations of
motion, suggests that such features can only be seen in a full
non-perturbative treatment of the dynamics.
An application of our results in a quantum-cosmological context has further
highlighted important questions which are usually not addressed in
minisuperspace constructions, related for instance to coarse graining and
stability. 

\section*{Acknowledgements}

This work was supported in part by NSF grants PHY-1417385 and PHY-1607414.


\end{document}